\documentclass[twocolumn]{aastex631}

\usepackage{siunitx}
\usepackage{physics}
\usepackage{svg}
\usepackage[caption=false]{subfig}
\usepackage{tikz,tikz-3dplot,yhmath}
\usetikzlibrary{angles,quotes}
\usepackage{xcolor}

\DeclareSIUnit\jansky{Jy}
\DeclareSIUnit\parsec{pc}
\DeclareSIUnit\year{yr}
\DeclareSIUnit\erg{erg}
\DeclareSIUnit\photon{photon}
\DeclareSIUnit\rad{rad}
\DeclareSIUnit\sr{sr}
\DeclareSIUnit\magnitude{mag}
\DeclareSIUnit\Msun{M_{\odot}}
\DeclareFontFamily{OMX}{yhex}{}
\DeclareFontShape{OMX}{yhex}{m}{n}{<->yhcmex10}{}
\DeclareSymbolFont{yhlargesymbols}{OMX}{yhex}{m}{n}
\DeclareMathAccent{\wideparen}{\mathord}{yhlargesymbols}{"F3}

\shorttitle{FRBs from rotation modulated magnetar starquakes}
\shortauthors{J-W Luo et al.}

\begin{document}
\title{Hyper-active repeating fast radio bursts from rotation modulated starquakes on magnetars}

\author[0000-0002-9642-9682]{Jia-Wei Luo}
\affiliation{College of Physics and Hebei Key Laboratory of Photophysics Research and Application, Hebei Normal University, Shijiazhuang, Hebei 050024, China}
\affiliation{Shijiazhuang Key Laboratory of Astronomy and Space Science, Hebei Normal University, Shijiazhuang, Hebei 050024, China}
\author[0000-0001-8065-4191]{Jia-Rui Niu}
\affiliation{National Astronomical Observatories, Chinese Academy of Sciences, Beijing 100101, China}
\author[0000-0001-9036-8543]{Wei-Yang Wang}
\affiliation{School of Astronomy and Space Science, University of Chinese Academy of Sciences, Beijing 100049, China} 
\author[0000-0002-8744-3546]{Yong-Kun Zhang}
\affiliation{National Astronomical Observatories, Chinese Academy of Sciences, Beijing 100101, China}
\affiliation{School of Astronomy and Space Science, University of Chinese Academy of Sciences, Beijing 100049, China}
\author[0000-0002-6423-6106]{De-Jiang Zhou}
\affiliation{National Astronomical Observatories, Chinese Academy of Sciences, Beijing 100101, China}
\affiliation{School of Astronomy and Space Science, University of Chinese Academy of Sciences, Beijing 100049, China} 
\author[0000-0002-5031-8098]{Heng Xu}
\affiliation{National Astronomical Observatories, Chinese Academy of Sciences, Beijing 100101, China}
\author[0000-0002-3386-7159]{Pei Wang}
\affiliation{National Astronomical Observatories, Chinese Academy of Sciences, Beijing 100101, China}
\affiliation{Institute for Frontiers in Astronomy and Astrophysics, Beijing Normal University,  Beijing 102206, China}
\author[0000-0001-6651-7799]{Chen-Hui Niu}
\affiliation{Central China Normal University, Wuhan 430079, China}
\author[0009-0000-1470-519X]{Zhen-Hui Zhang}
\affiliation{College of Physics and Hebei Key Laboratory of Photophysics Research and Application, Hebei Normal University, Shijiazhuang, Hebei 050024, China}
\affiliation{Shijiazhuang Key Laboratory of Astronomy and Space Science, Hebei Normal University, Shijiazhuang, Hebei 050024, China}
\author[0000-0003-2413-9587]{Shuai Zhang}
\affiliation{College of Physics and Hebei Key Laboratory of Photophysics Research and Application, Hebei Normal University, Shijiazhuang, Hebei 050024, China}
\affiliation{Shijiazhuang Key Laboratory of Astronomy and Space Science, Hebei Normal University, Shijiazhuang, Hebei 050024, China}
\author[0000-0002-6540-2372]{Ce Cai}
\affiliation{College of Physics and Hebei Key Laboratory of Photophysics Research and Application, Hebei Normal University, Shijiazhuang, Hebei 050024, China}
\affiliation{Shijiazhuang Key Laboratory of Astronomy and Space Science, Hebei Normal University, Shijiazhuang, Hebei 050024, China}
\author[0000-0002-9274-3092]{Jin-Lin Han}
\affiliation{National Astronomical Observatories, Chinese Academy of Sciences, Beijing 100101, China}
\author[0000-0003-3010-7661]{Di Li}
\affiliation{New Cornerstone Science Laboratory, Department of Astronomy, Tsinghua University, Beijing 100084, China}
\affiliation{National Astronomical Observatories, Chinese Academy of Sciences, Beijing 100101, China}
\author[0000-0002-1435-0883]{Ke-Jia Lee}
\affiliation{Department of Astronomy, Peking University, Beijing 100871, China}
\affiliation{Kavli Institute for Astronomy and Astrophysics, Peking University, Beijing 100871, China}
\author[0000-0001-5105-4058]{Wei-Wei Zhu}
\affiliation{National Astronomical Observatories, Chinese Academy of Sciences, Beijing 100101, China}
\affiliation{School of Astronomy and Space Science, University of Chinese Academy of Sciences, Beijing 100049, China}
\author[0000-0002-9725-2524]{Bing Zhang}
\affiliation{Nevada Center for Astrophysics, University of Nevada, Las Vegas, NV 89154, USA}
\affiliation{Department of Physics and Astronomy, University of Nevada, Las Vegas, NV 89154, USA}

\correspondingauthor{Jia-Wei Luo}
\email{ljw@hebtu.edu.cn},
\correspondingauthor{Wei-Wei Zhu}
\email{zhuww@nao.cas.cn},
\correspondingauthor{Bing Zhang}
\email{bing.zhang@unlv.edu}

\begin{abstract}
     The non-detection of periodicity related to rotation challenges magnetar models for fast radio bursts (FRBs) with FRB emission from close to the magnetar surface. Moreover, a bimodal distribution of the burst waiting times is widely observed in hyper-active FRBs, a significant deviation from the exponential distribution expected from stationary Poisson processes. By combining the epidemic-type aftershock sequence (ETAS) earthquake model and the rotating vector model (RVM) involving the rotation of the magnetar and orientations of the spin and magnetic axes, we find that starquake events modulated by the rotation of FRB-emitting magnetar can explain the bimodal distribution of FRB waiting times, as well as the non-detection of periodicity in hyper-active repeating FRBs. We analyze data from multiple FRB sources, demonstrating that differences in waiting time distributions and, to some extent, observed energies can be explained by varying parameters related to geometric properties of the magnetar FRB emission and starquake dynamics. Our results show that the assumption that all FRBs are repeaters is compatible with our model. Notably, we find that hyper-active repeaters tend to have small magnetic inclination angles in order to hide their periodicity. We also show that our model can reproduce the waiting time distribution of a pulsar phase of the galactic magnetar SGR J1935+2154 with a larger inclination angle than the hyper-active repeaters, which could explain the detection of spin period and the relatively low observed energy for FRBs from the magnetar. The spin periods of hyper-active repeaters are not well constrained, but most likely fall in the valley region between the two peaks of the waiting time distributions. 
\end{abstract}

\section{Introduction}
\label{sec:introduction}

Fast Radio Bursts (FRBs) are enigmatic and highly energetic astronomical phenomena characterized by transient radio pulses lasting for extremely brief durations, typically from milliseconds down to nanoseconds \citep{nimmo2022BurstTimescalesLuminosities}. Discovered in 2007 by \cite{lorimer2007BrightMillisecondRadio}, these bursts are notable for their extreme brightness and short duration, which make them both challenging and intriguing for detection and analysis. Interestingly, some FRBs are found to repeat \citep{spitler2016RepeatingFastRadio}, while most others have yet to show repetition \citep{chime/frbcollaboration2021FirstCHIMEFRB}. Despite active research, the precise origin and mechanism behind FRBs remain one of the most compelling mysteries in modern astrophysics. Numerous theories have been proposed with varying degrees of success, but none are able to account for all of the observed FRB phenomena (for a comprehensive review of FRB theories, see \citealt{zhang2023PhysicsFastRadio}). 

Currently, the most compelling source model for FRBs is the magnetar model \citep{popov2007HyperflaresSGRsEngine,thornton2013PopulationFastRadio,lyubarsky2014ModelFastExtragalactic,kulkarni2014GIANTSPARKSCOSMOLOGICAL,katz2016HOWSOFTGAMMA,kumar2017FastRadioBurst,metzger2017MillisecondMagnetarBirth,beloborodov2017FlaringMagnetarFRB,yang2018BunchingCoherentCurvature,metzger2019FastRadioBursts,beloborodov2020BlastWavesMagnetar,margalit2020ConstraintsEnginesFast,wadiasingh2020FastRadioBurst,lyubarsky2020FastRadioBursts,yang2020PairSeparationParallel,lu2020UnifiedPictureGalactic,yang2021FastRadioBursts,wang2022MagnetosphericCurvatureRadiation,zhang2022CoherentInverseCompton,beniamini2023HybridPulsarMagnetar,qu2024CoherentInverseCompton}. A crucial piece of evidence for the magnetar model is the detection of FRB 20200428, which is the first FRB to be conclusively associated with a magnetar, SGR J1935+2154 within our own Galaxy \citep{chime/frbcollaboration2020BrightMilliseconddurationRadio,bochenek2020FastRadioBurst}.

However, the magnetar models are not without issues. Mounting observational evidence suggests that the emission likely originates within the magnetospheres \citep{luo2020DiversePolarizationAngle,xu2022FastRadioBurst,zhang2023FASTObservationsFRB,niu2024SuddenPolarizationAngle,nimmo2025MagnetosphericOriginFast,mckinven2025PulsarlikePolarizationAngle,jiang2025NinetyPercentCircular}. Within such a scenario, clear periodicity is expected (as observed in radio pulsars). However, although many searches have been conducted, credible periodicity of FRBs in the realm of magnetar spin period was not found in most FRBs \citep[e.g.][]{katz2022AbsencePeriodicityRepeating,niu2022FASTObservationsExtremely,du2024ThoroughSearchShorttimescale}, with the only exception of the 0.286s period detected in FRB 20191221A \citep{thechime/frbcollaboration2022SubsecondPeriodicityFast}. Longer periodicities are observed in some FRBs, such as the 16.35-day periodicity detected in FRB 20180916B \citep{thechime/frbcollaboration2020PeriodicActivityFast} or the $\sim160$ day \citep{rajwade2020PossiblePeriodicActivity,cruces2020RepeatingBehaviourFRB} and 4.605-day \citep{li2024CandidatePeriod4605} periodicities detected in FRB 20121102A, but it would require extreme conditions for magnetars to have such slow spin periods \citep{beniamini2020PeriodicityRecurrentFast,xu2021PeriodicActivitiesRepeating}. Indeed, most studies associate these longer periodicities with the orbital period in binary systems \citep[e.g][]{ioka2020BinaryCombModel,lyutikov2020FRBPeriodicityMild,li2021PeriodicActivitiesRepeating,wada2021BinaryCombModels,wang2022RepeatingFastRadio,zhao2023RotationMeasureVariations,braga2025FRB20121102AMonitoring,li2025StructureFunctionsRotation} or precession of magnetars \citep[e.g.][]{beniamini2020PeriodicityRecurrentFast,yang2020OrbitinducedSpinPrecession,levin2020PrecessingFlaringMagnetar,feng2024PeriodicActivitiesFast}. The non-detection of periodicity associated with the spin period of the FRB-emitting magnetars severely challenges the magnetar model for FRBs. Moreover, some hyper-active repeating FRBs have been observed to repeat thousands of times in months or even days \cite[e.g.][]{li2021BimodalBurstEnergy,xu2022FastRadioBurst,zhang2023FASTObservationsFRB}, making it difficult to explain the energy budget of hyper-active FRBs solely based on the magnetic energy stored in the magnetosphere of magnetars.

On the other hand, the waiting time distributions of many hyper-active repeating FRBs are found to be bimodal \citep{li2021BimodalBurstEnergy,xu2022FastRadioBurst,zhang2022FASTObservationsExtremely,zhang2023FASTObservationsFRB}. This bimodality appears to be a common feature of hyper-active repeating FRBs, yet its physical origin remains unknown. Some efforts have been made to explain the reason behind this bimodality, but no solid conclusion has been reached. \cite{li2021BimodalBurstEnergy} postulates that the longer peak in the waiting time distribution is associated with the general event rate of bursts, while the shorter peak is related to the substructure of individual bursts. \cite{xiao2024PropagationFastRadio} suggest that the observed waiting time distribution of hyper-active FRBs can form from propagation effects in the magnetosphere, albeit a bimodal distribution is needed to begin with.

Some studies point out that the temporal behavior of hyper-active FRBs is similar to that of earthquakes \citep{totani2023FastRadioBursts, wang2023RepeatingFastRadio, tsuzuki2024SimilarityEarthquakesAgain, gao2024ComparativeAnalysisScaleinvariant, sang2024QuantifyingRandomnessScale}, which remarkably also display bimodality in their waiting time distributions \citep[e.g.][]{touati2009OriginNonuniversalityEarthquake,talbi2010MixedModelEarthquake}. Concomitantly, starquakes have been theorized as a possible triggering mechanism for FRBs \citep{wang2018FRB121102Starquakeinduced,wadiasingh2019RepeatingFastRadio,suvorov2019YoungMagnetarsFracturing,yuan2020PlasmoidEjectionAlfven,wadiasingh2020FastRadioBursta,li2021PeriodicActivitiesRepeating,yang2021FastRadioBursts,zhang2022CoherentInverseCompton,li2022RepeatingFastRadio,lander2023GameLifeMagnetar,qu2024CoherentInverseCompton,xie2024FindingParticularityActive,wang2024EnergyBudgetStarquakeinduced,bilous2025ActivityTransitionFRB,yamasaki2025TimeFrequencyCorrelation,wu2025UniversalBreakEnergy}. Furthermore, glitch events were observed around FRB events of SGR J1935+2154 \citep{ge2022GiantGlitchMagnetar,younes2023MagnetarSpindownGlitch,hu2024RapidSpinChanges}, suggesting crust interactions might be one of the key ingredients for triggering an FRB.

In this paper, we attempt to explain the bimodal distributions of FRB waiting time with a rotation modulated magnetar starquake model integrating the epidemic-type aftershock sequence (ETAS) earthquake model and the rotating vector model (RVM) involving the geometry of FRB emission. This paper is organized as follows: In Section \ref{sec:data}, we introduce the FRB data we utilize. In Section \ref{sec:methods}, we elucidate the details of our model. In Section \ref{sec:results}, we present the fitting results and reproduce the bimodal distributions of the FRB waiting time with our model. In Section \ref{sec:discussion}, we extend some discussion on our model and how it might lead to a unified model for repeating and non-repeating FRBs. In Section \ref{sec:conclusions}, we put forward our conclusions.

\section{Data}
\label{sec:data}

In this study, we utilize 5 FRB datasets obtained by the Five-hundred-meter Aperture Spherical Telescope (FAST, \citealt{nan2011FiveHundredApertureSpherical}): 1652 bursts detected from FRB 20121102A over 59.5 hours across 47 days Between August 29 and October 29, 2019\footnote{\url{https://www.scidb.cn/en/detail?dataSetId=f172ff40142c4100855724b80a085deb}} \citep{li2021BimodalBurstEnergy}; 1863 bursts detected from FRB 20201124A over 82 hours across 54 days between April 1 and June 11, 2021\footnote{\url{https://psr.pku.edu.cn/index.php/publications/frb20201124a/}} \citep{xu2022FastRadioBurst}; 881 bursts detected from FRB 20201124A over 4 hours across 4 days in a very active episode between September 25 -- 28, 2021\footnote{\url{https://www.scidb.cn/en/detail?dataSetId=66cf4a4f2f334739bfa4223d381b936f}} \citep{zhou2022FASTObservationsExtremely}; 1076 bursts detected from FRB 20220912A over 8.67 hours across 55 days between October 28 and December 22, 2022\footnote{\url{https://www.scidb.cn/en/detail?dataSetId=18a982901dd244f6b236bab445095e57}} \citep{zhang2023FASTObservationsFRB}; 219 bursts FRB detected from FRB 20190520B across a long monitoring campaign between 2019 and 2021\footnote{\url{https://www.scidb.cn/en/detail?dataSetId=d0d823e5c5ba4c32ae32fc37c931472a}} \citep{zhang2024ArrivalTimeEnergy}. All 5 datasets are publicly available in the corresponding references. To distinguish between the two epochs from FRB 20201124A, we refer to the April -- June dataset as 20201124A(A) and the September dataset as 20201124A(B), respectively.

We also used a dataset from a radio pulsar phase of the Galactic magnetar SGR J1935+2154, which consists of 796 pulses observed over 6.5 hours across 13 days between 9 October 2020 -- 30 October 2020\footnote{\url{http://groups.bao.ac.cn/psr/xsdt/202206/t20220607_704493.html}} \citep{zhu2023RadioPulsarPhase}. Note that radio pulses from this pulsar phase are significantly different from the FRBs from this source, with a much higher event rate and lower energies. We choose to use the pulsar phase data because it has a comparable event rate with the hyper-active FRBs and a similar bimodal waiting time distribution.

The burst energies are calculated with frequency bandwidth:
\begin{equation}
    E=4\pi D_{\rm L}^2 F \Delta\nu /(1+z),
\end{equation}
where $F$ is the fluence of the FRB, and $\Delta\nu$ is frequency bandwidth. The adoption of $\Delta\nu$ over the central frequency $\nu_0$ is preferred for repeater bursts because of their narrow spectra observable within the telescope frequency band \citep{zhang2023PhysicsFastRadio}. For FRB 20121102A $z=0.19273$ \citep{tendulkar2017HostGalaxyRedshift}, for FRB 20201124A $z=0.09795$ \citep{xu2022FastRadioBurst}, for FRB 20220912A $z=0.0771$ \citep{ravi2023DeepSynopticArray}, for FRB 20190520B $z=0.241$ \citep{niu2022RepeatingFastRadio}, for SGR J1935+2154 $D_L= \SI{6.6}{\kilo\parsec}$ \citep{zhou2020RevisitingDistanceEnvironment}. The luminosity distances are calculated with cosmology parameters measured by Planck \citep{aghanim2020Planck2018Results}: $H_0=\SI{67.4}{\kilo\meter\per\second\per\mega\parsec}$, $\Omega_m=0.315$, $\Omega_b=0.0493$ and $\Omega_{\Lambda}=0.685$.

Waiting time is calculated as the difference in arrival times of each burst and its preceding burst:
\begin{equation}
\tau_i = t_i-t_{i-1}.
\end{equation}

To avoid the results being affected by the observation cadence, we only calculate waiting times between bursts in the same observing session. 

There are a few special outlier data points that we have excluded from the data: 
\begin{itemize}
\item The waiting time between bursts \#101 and \#102 of the FRB 20220912A dataset is shorter than the data precision. Therefore, we do not include this waiting time.
\item
The waiting time between bursts \#551 and \#552 of the FRB 20201124A(B) dataset is much shorter than other waiting times in the same dataset. While this waiting time may represent genuine physical separations, the corresponding histogram bins would exhibit disproportionately high uncertainties due to potential detection ambiguities in distinguishing temporally proximate bursts. To avoid this outlier affecting our results, we do not include this waiting time.
\item
The energies of bursts \#677 of the FRB 20121102A dataset and \#452 of the FRB 20220912A dataset are much lower than other energies in the same dataset. To avoid these outliers affecting our results, we do not include these energies.
\item
We do not include the first 4 bursts of FRB 20190520B and the first burst of SGR J1935+2154 as they are too far from other observations.
\end{itemize}

\section{Methods}
\label{sec:methods}

\subsection{The epidemic-type aftershock sequence earthquake model}

The epidemic-type aftershock sequence (ETAS) model \citep{ogata1988StatisticalModelsEarthquake} is shown to be one of the best-performing models in reproducing the event rates of earthquakes \citep[e.g.][]{nandan2019ForecastingRatesFuture}. The ETAS model is based on a Hawkes process (for a comprehensive review of Hawkes processes, see \citealt{laub2021ElementsHawkesProcesses}) where the earthquake event rate follows an inhomogeneous Poisson process consisting of two components, a stationary Poisson term representing background activity (mainshocks) and the sum of triggering rates of the history of earthquakes prior to the current time representing triggered activities (aftershocks):
\begin{equation}
\lambda_{\theta}(t,m|\mathcal{H}) = f(m)\qty[\mu + \sum_{i:t_i<t} g(\Delta t_i,m_i)],
\end{equation}
where $\lambda_{\theta}$ is the earthquake event rate at time $t$ and magnitude $m$, $\theta$ is the parameters for the ETAS model, $\mathcal{H}$ represents the history of earthquake events, $f(m)$ is the probability distribution of earthquake magnitudes, $\mu$ is the constant background rate, $t_i$ is the time of the $i$th event prior to the current time $t$, $g(\Delta t_i,m_i)$ is the activation function representing the expected aftershock rate at time $t$ triggered by earthquake $i$ and is a function of the difference in time $\Delta t_i=t-t_i$ and the magnitude of the triggering earthquake $m_i$.

Note that we distinguish between mainshocks and aftershocks in neither the summation of the activation function nor the magnitude distribution function in the above equation. In the ETAS model, there is no physical difference between mainshocks and aftershocks, the distinction is only made statistically. Therefore, aftershocks can produce second and higher generation aftershocks, and there is no requirement that the magnitude of aftershocks should be smaller than their triggering mainshocks. As a result, one major earthquake can produce many generations of aftershocks, forming an earthquake cluster.

Studies show the ETAS model can reproduce bimodal distribution in waiting time as observed in many real-world earthquake data \citep[e.g.][]{touati2009OriginNonuniversalityEarthquake,talbi2010MixedModelEarthquake}, with the longer peak in waiting time corresponding to background mainshocks and the shorter peak corresponding to the triggered aftershocks.

It would be a misconception to assume that the bimodal distribution of FRB waiting time is a product of two physically distinct random processes, as the addition of two independent stationary Poisson processes will only result in a new stationary Poisson process with a higher event rate, thus a bimodal distribution in waiting time cannot be produced. Therefore, models similar to the ETAS model that involve two correlated components are needed. Furthermore, if the two peaks of the waiting time of FRBs correspond to two physically distinct processes, bursts having short waiting times with adjacent bursts and those having long waiting times with adjacent bursts would exhibit different distributions in the observational parameters. However, such a difference has not been reported currently. Our model, on the other hand, is able to explain the bimodal waiting time distribution with a single physical process.

The ETAS model uses the Gutenberg–Richter law \citep{gutenberg1944FrequencyEarthquakesCalifornia} as the distribution of earthquake magnitudes:
\begin{equation}
f(m)=\beta_e \exp\qty[-\beta_e (m-m_c)],
\end{equation}
where $\beta_e$ is a constant index, and $m_c$ is a reference magnitude. Note that we refer to the $\beta$ parameter in the Gutenberg–Richter law as $\beta_e$ to distinguish it from the impact parameter $\beta$ we later introduce.

Since the magnitude of earthquakes is defined with the logarithmic energy scale similar to the magnitude defined in astronomy (but note that higher earthquake magnitude corresponds to higher energy, contrary to the astronomical magnitude), an exponential distribution of magnitude corresponds to a power-law distribution in earthquake energy with a power-law index of $\beta_e$, again similar to most luminosity functions in astronomy. In this study, we simply define the magnitude to be the base 10 logarithm of the isotropic FRB energy, $m=\log E$. One magnitude defined in seismology does not correspond to one order of magnitude in energy. However, different magnitude--energy factors only result in a scale difference between the earthquake parameters and FRB parameters. 

As for the activation function, the modified Omori-Utsu law \citep{utsu1961StatisticalStudyOccurrence} and the Utsu aftershock productivity law \citep{utsu1970AftershocksEarthquakeStatistics} are employed:
\begin{equation}
\label{eq:activation_function}
g(\Delta t,m) = \frac{K e^{a (m-m_c)}}{(\Delta t + c)^p},
\end{equation}
where $m_c$ is the minimum magnitude for earthquakes, which is needed for the simulation to converge (we use the same $m_c$ in magnitude distribution and activation function for convenience), and $K$, $a$, $c$, $p$ are constants. The aftershock productivity parameter $K$ represents the base rate of aftershock generation per mainshock. The magnitude sensitivity parameter $a$ scales how much a unit increase in mainshock magnitude boosts aftershock productivity. The time offset parameter $c$ is a characteristic time that shifts the time decay law to prevent infinite triggering rates immediately after a mainshock. The time decay exponent (also known as the Omori exponent) $p$ controls how rapidly the aftershock rate declines. Together, the function $g(\Delta t,m)$ represents a delayed power-law decay in the triggering rate as the time difference increases and exponential growth in the triggering rate as the magnitude of the triggering event increases.

The ETAS model was later modified to include locations of earthquakes to make more accurate predictions \citep{ogata1998SpaceTimePointProcessModels,ogata2006SpaceTimeETAS}. However, since we do not have location data of FRBs on magnetars, we opt to use the simpler version with fewer free parameters.

\subsection{ETAS simulations}

We follow the methods described in \cite{mizrahi2021EffectDeclusteringSize,mizrahi2021EmbracingDataIncompleteness} to simulate earthquake events based on the ETAS model.

The primary simulation time is defined as the difference in time between the start of the first observation and the end of the last observation of an FRB. However, because the event rate at the start of the simulation did not include earlier triggering events, we include an extra burn-in time equal to half of the primary simulation time, the events in which are discarded later. After defining the simulation time range, we first simulate background events. For background events, their time distribution and magnitude distribution are independent, thus can be generated separately. 

To simulate the event time of background events, we first calculate the number of background events by drawing a random number from a Poisson distribution with the expectation of $\mu T$, where $T$ is the total duration of the simulation. Then, the times of background events are drawn from a uniform distribution between the start time $T_i$ and end time $T_n$ of the simulation. Finally, the magnitudes of the background events are drawn from $f(m)$.

We can then calculate the number of expected aftershocks of each mainshock as a function of mainshock magnitude, under the assumption that the total simulation time is infinite:
\begin{equation}
    n_{AS}(m) = \int_0^{\infty} g(\Delta t,m)\, d\Delta t=\frac{Kc^{-(p-1)}e^{a(m-m_c)}}{p-1}.
\end{equation}

From the number of expected aftershocks, we can define an important parameter, the branching ratio:
\begin{equation}
    \eta = \int_{m_c}^{\infty} f(m)n_{AS}(m)\, dm=\frac{K\beta_e c^{-(p-1)}}{(\beta_e-a)(p-1)}.
\end{equation}

Note that $\beta_e>a$ is required for the above integral to converge. The branching ratio indicates the average number of direct (first-generation) aftershocks of any earthquake. Under the assumption that the total simulation time $T$ is infinite, $\eta<1$ is required for the simulation to converge. 

We assign the actual number of aftershocks to each mainshock by drawing samples from Poisson distributions with expectations set to their expected number of aftershocks $n_{AS}$. The event times and magnitudes of aftershocks are subsequently drawn from $g(\Delta t_i, m_i)$ and $f(m)$ respectively. Aftershocks with times outside of the simulation time range are dropped.

After we generate the first generation of aftershocks, we now treat them as ``mainshocks" for the next generation of aftershocks. The aftershocks are generated iteratively until the total actual number of aftershocks is zero for the current generation. The full catalog of earthquake events is the combination of the background events and all generations of the triggered events. Finally, the simulated events are filtered according to the observation log to only include those that lie within the observation cadence to simulate the gaps between observations. However, note that for simulations with the fitted parameters in this paper, this filtering does not change the results significantly.

For the courtesy to the readers, we rewrite the equations above replacing magnitude $m$ with the logarithm of isotropic energy $\log E$.

\begin{equation}
\lambda_{\theta}(t,\log E|\mathcal{H}) = f(\log E)\qty[\mu + \sum_{i:t_i<t} g(\Delta t_i,\log E_i)],
\end{equation}
\begin{equation}
f(\log E)=\beta_e \exp\qty[-\beta_e (\log E-\log E_c)],
\end{equation}
\begin{equation}
g(\Delta t,\log E) = \frac{K e^{a (\log E-\log E_c)}}{(\Delta t + c)^p},
\end{equation}
\begin{equation}
    n_{AS}(\log E) = \int_0^{\infty} g(\Delta t,\log E)\, d\Delta t=\frac{Kc^{-(p-1)}e^{a(\log E-\log E_c)}}{p-1},
\end{equation}
where $\log E_i$ is the logarithm of isotropic energy of the $i$th event, $\log E_c$ represents the reference energy.

\subsection{The magnetar rotating vector model}

The rotating vector model (RVM) was first proposed to explain the polarization properties of pulsars \citep{radhakrishnan1969MagneticPolesPolarization}. The polarization angle (PA) evolution of some FRBs is found to resemble RVM predictions \citep{pandhi2024PolarizationProperties128,mckinven2025PulsarlikePolarizationAngle}, even though most of them either show flat PA \citep{michilli2018ExtremeMagnetoionicEnvironment,jiang2022FASTObservationsExtremely} or sometimes PA jumps \citep{niu2024SuddenPolarizationAngle,jiang2025NinetyPercentCircular}. 

Judging from the complex polarization properties of FRBs \citep[e.g.][]{zhang2023FASTObservationsFRB,pandhi2024PolarizationProperties128}, it could be possible that the FRB-emitting magnetars do not have simple dipole fields \citep[e.g.][]{zhu2023RadioPulsarPhase} and the extremely high luminosities of FRBs might significantly distort the local magnetic field \citep[e.g.][]{ioka2020FastRadioBurst,qu2022TransparencyFastRadio}. As a result, RVM alone cannot fully explain the PA evolution of FRBs. In this paper, we adopt the RVM solely to model the FRB emission geometry for reproducing the waiting time and energy distributions, while a comprehensive analysis of polarization properties within our framework will be addressed in future studies.

\begin{figure}
    \centering
    \includegraphics[width=0.49\textwidth]{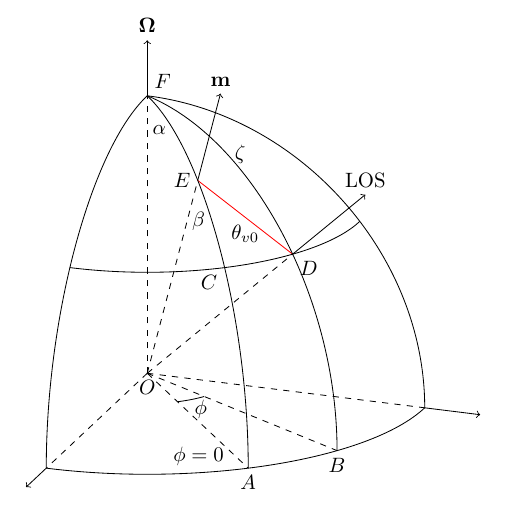}
    \caption{An illustration of the geometry of the rotating vector model of FRB-emitting magnetars used in this paper. The central angle corresponding to the red arc $\wideparen{ED}$ between the magnetic axis and the line of sight is the viewing angle $\theta_{v0}$ excluding the effect of $\rho_{max}$.}
    \label{fig:rvm}
\end{figure}

The RVM used in this paper is shown in Fig. \ref{fig:rvm}. All FRBs are assumed to be roughly beamed toward the direction of the magnetic axis $\mathbf{m}$, and we allow the FRB directions to randomly and isotropically deviate from the magnetic axis within a set range of $\rho_{max}$. The inclination angle $\alpha$ is the angular distance between the spin axis $\mathbf{\Omega}$ and the magnetic axis, the impact parameter $\beta$ is the minimum angular distance between the magnetic axis and the line of sight, $\beta>0$ if the line of sight and the spin axis are on opposite sides of the magnetic pole, otherwise $\beta<0$, and the line of sight colatitude is $\zeta=\alpha+\beta$. The zero of spin longitude $\phi$ is defined as the meridian through the magnetic axis. As the FRB-emitting magnetar rotates, the angular distance between the line of sight and the magnetic axis, viewing angle $\theta_{v0}$, changes corresponding to $\phi$. 

One can write the equation for $\theta_{v0}$ by invoking the cosine law:
\begin{equation}
    \theta_{v0}=\arccos\qty(\cos\zeta\cos\alpha+\sin\zeta\sin\alpha\cos\phi).
    \label{eq:theta_v0}
\end{equation}

\begin{figure}
    \centering
    \begin{tikzpicture}
    \coordinate  (v) at (3,0);
    \coordinate  (m) at (-2,-1);
    \coordinate  (rho) at (-1,2);
    
    \draw[thick] (v) -- (m) node[midway, text=black,anchor=north]{$\theta_{v0}$};
    \draw[thick] (v) -- (rho) node[midway, text=black,anchor=south]{$\theta_v$};
    \draw[thick] (m) -- (rho) node[midway, text=black,anchor=east]{$\rho$};
    
    \filldraw[] (v) circle (2pt) node[anchor=west]{LOS};
    \filldraw[] (m) circle (2pt) node[anchor=east]{$\mathbf{m}$};
    \filldraw[] (rho) circle (2pt) node[anchor=south]{FRB beam};

    \draw[thick] (v) -- (m) -- (rho) pic["$\theta_{\rho}$", draw, angle radius=1cm] {angle=v--m--rho};
    
    \end{tikzpicture}
    \caption{An illustration showing the orientations of $\rho$, $\theta_{\rho}$, $\theta_{v0}$ and $\theta_v$. Note that the angles are on a sphere and this is only a schematic diagram.}
    \label{fig:rho}
\end{figure}
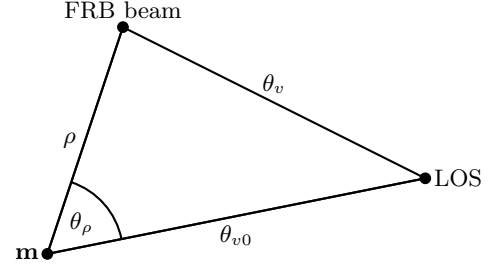

To include the effect of $\rho_{max}$, we further introduce two quantities as shown in Fig. \ref{fig:rho}. The first quantity is the angular distance between the FRB beam and the magnetic axis, $\rho\in[0,\rho_{max}]$. To define the second quantity $\theta_{\rho}$, we first connect the FRB beam direction and the magnetic axis, then $\theta_{\rho}$ is the angle between this line and the line between the magnetic axis and the line of sight $\wideparen{ED}$. $\theta_{\rho}=0$ when the two lines are in the same direction.

Finally, we can obtain the equation for $\theta_v$ considering $\rho$ and $\theta_{\rho}$ by invoking the cosine law again:
\begin{equation}
\theta_v=\arccos\qty(\cos\theta_{v0}\cos\rho+\sin\theta_{v0}\sin\rho\cos\theta_{\rho}),
\end{equation}
where $\theta_{v0}$ is the viewing angle when $\rho=0$ as calculated with Eq. \ref{eq:theta_v0}.

\subsection{Beaming structure}

The observed energy of an FRB is its intrinsic energy multiplied by the beaming structure function. In this paper, we assume that the beaming structure has a rotational symmetry around the FRB beaming axis, thus the observed energy $E_o$ is only a function of the intrinsic energy $E_i$ and viewing angle $\theta_v$.

We consider the Doppler factor defined in \cite{zhang2021SlowRadioBursts,chen2023FRBSRBXRB}:
\begin{equation}
    \mathcal{D}(\theta_v)=\frac{1}{\Gamma(1-\beta_d\cos\theta_v)},
\end{equation}
where $\Gamma$ is the Lorentz factor, $\beta_d$ is the dimensionless speed. Note that we did not invoke an ``on-axis" region as in the above-mentioned references, as it is already included by $\rho_{max}$.

\begin{figure}
    \centering
    \includegraphics[width=0.49\textwidth]{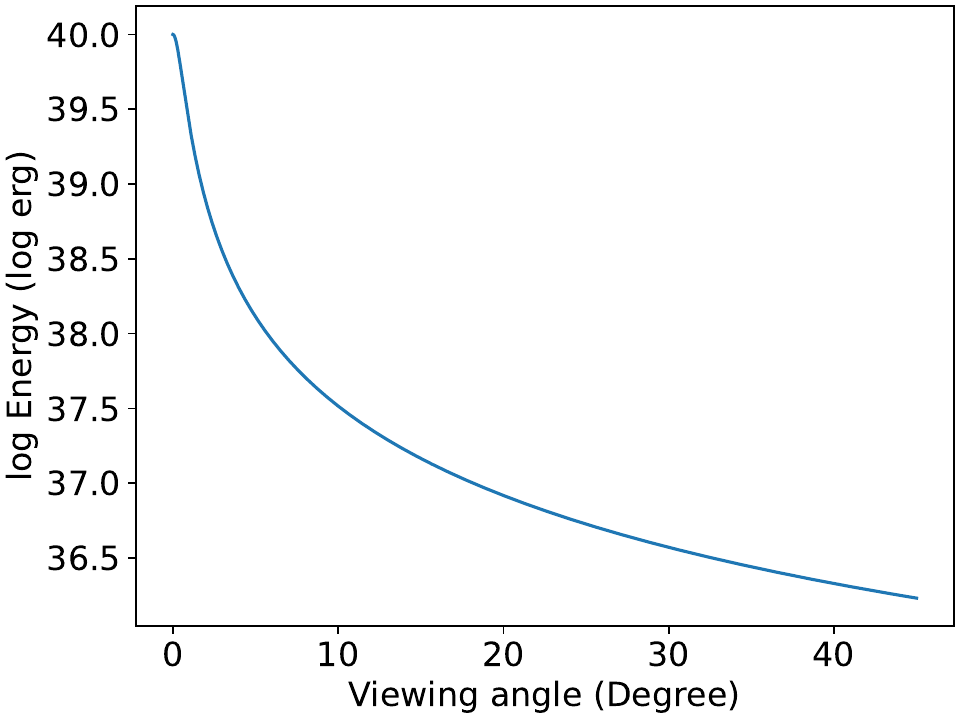}
    \caption{Angular energy distribution with $\Gamma=100$ and $E_i=10^{40} \si{\erg}$.}
    \label{fig:energy_angle}
\end{figure}

The Doppler factor then connects the observed energy $E$ and the intrinsic energy $E'$ at $\theta_v$ with (e.g. \cite{zhang2018PhysicsGammaRayBursts}):
\begin{equation}
    E = \mathcal{D}E'.
\end{equation}

To simplify the simulation, we normalize the Doppler factor and consider the relationship between the observed isotropic energy $E_o$ and the intrinsic isotropic energy $E_i$:
\begin{equation}
    E_o = \frac{\mathcal{D}}{\mathcal{D}_0}E_i,
\end{equation}
where $D_0=\frac{1}{\Gamma(1-\beta_d)}$ is the Doppler factor at $\theta_v=0$.
An example of the beaming structure with $\Gamma=100$ and $E_i=10^{40} \si{\erg}$ is shown in Fig. \ref{fig:energy_angle}.

According to this picture, a larger $\theta_v$ also predicts a longer duration (width $w$) \citep{zhang2021SlowRadioBursts}, i.e.
\begin{equation}
    w_o = \frac{\mathcal{D}_0}{\mathcal{D}}w_i. 
\end{equation}
\cite{brown2023ValidatingSubburstSlope} found an inverse relationship between sub-burst duration and frequency, consistent with the two parameters being governed by the common Doppler factor. 

We set the detection probability of FRBs as a logistic function defined on the basis of the observed distribution of the burst energy. The detection probabilities are shown in Appendix Figure \ref{fig:dect_prob}.

\subsection{MCMC}
 
We use the Markov Chain Monte Carlo (MCMC) method implemented by the \textsc{Python} package \textsc{emcee} \citep{foreman-mackey2013EmceeMCMCHammer} to fit the parameters in our model. The prior distributions are uniform, and the ranges are listed in Table \ref{table:prior_range}. Since the ranges of many parameters, especially those related to starquakes, are theoretically poorly predicted, we adopted wide enough ranges for our MCMC parameter searches. 
We also require $\beta_e>a$ to make sure that the simulation will converge. The branching ratio $\eta$ is constrained to be $\eta<0.8$. Although simulations can converge when $\eta\in[0.8,1)$, Larger $\eta$ values produce excessive aftershocks, leading to unrealistically high probability density at short waiting times in the distribution, which contradicts observational constraints. The high number of aftershocks would also significantly decelerate computation.

\begin{table}
\centering
\hspace*{-1.2cm}
\addtolength{\tabcolsep}{-3pt}
\begin{tabular}{lll}
    \hline
    Name & Unit & Range \\
    \hline
    $\alpha$ & $\si{\degree}$ & 0 -- 60\\
    $\beta$ & $\si{\degree}$ & $-\alpha$ -- 60\\
    $\log P$ & $\log \si{\s}$ & Min -- max observed waiting time\\
    $\Gamma$ & --- & 50 -- 1000\\
    $\rho_{max}$ & $\si{\degree}$ & 0 -- 60\\
    $\log \mu$ & $\log \si{\per\s}$ & -3 -- 5\\
    $\log K$ & --- & -5 -- 5\\
    $a$ & $\log \si{\per\erg}$ & -3 -- 3\\
    $\log c$ & $\log \si{\s}$& -5 -- 1\\
    $p$ & --- & 1.1 -- 9\\
    $\beta_e$ & --- & 1 -- 10\\
    $m_c$ & $\log \si{\erg}$ & 30 -- 45 (20 -- 30 for SGR J1935+2154)\\
    \hline
\end{tabular}
\addtolength{\tabcolsep}{3pt}
\caption{List of parameters and ranges for the MCMC. $\log P$ stands for base 10 logarithm for the spin period.}
\label{table:prior_range}
\end{table}

The likelihood function consists of three parts:
\begin{equation}
\mathcal{L} = \mathcal{L}_{wt} \cdot \mathcal{L}_{E} \cdot \mathcal{L}_{p},
\end{equation}
where $\mathcal{L}_{wt}$ and $\mathcal{L}_{E}$ are calculated by comparing the simulated waiting time and energy histogram distributions with the observed ones, and $\mathcal{L}_{p}$ represents the non-detection of periodicity.

The two histogram distribution likelihoods are calculated using the likelihood function defined by \cite{tremmel2013MODELINGREDSHIFTEVOLUTION}:
\begin{equation}
    \mathcal{L}_{histogram} = \prod_i \frac{\Gamma(\frac{1}{2}+d_i+n_i)}{2^{\frac{1}{2}+d_i+n_i}\Gamma(d_i+1)\Gamma(\frac{1}{2}+n_i)},
\end{equation}
where $d_i$ is the histogram for the observed data, $n_i$ is the is the histogram for simulated distribution, $\Gamma$ denotes the Gamma function here. We set the bin width $v_i=1$ in the original equation for simplicity as we used uniform bins.

We test the periodicity at the frequency corresponding to the spin period in our simulations with Lomb-Scargle periodogram implemented in \textsc{Astropy} \citep{theastropycollaboration2022AstropyProjectSustaining} and the false alarm level defined by \cite{baluev2008AssessingStatisticalSignificance}. Periodicity must not be detected at above $3\sigma$ level because deep searches of periodicity have constantly led to null results \citep{niu2022FASTObservationsExtremely,xu2022FastRadioBurst}. We adopt a threshold of 0.00135, which corresponds to the one-tailed significance level of $3\sigma$ in a standard normal distribution (also known as z-score). $\mathcal{L}_{p}=1$ if the calculated false alarm level is greater than $0.00135$, otherwise $\mathcal{L}_{p}=0$, rejecting the entire parameter combination. The periodicity requirement described above is not included for SGR J1935+2154, since its periodicity is clearly observed. 

A glossary of symbols used in this paper is listed in Appendix Table \ref{table:glossary}.

\section{Results}
\label{sec:results}

\subsection{Comparison of fitted parameters}
\label{subsec:parameters}

We apply the MCMC fitting to the data samples we have collected. The best-fit and 16th and 84th percentile values for the parameters are listed in Table \ref{table:fit_parameter}. The details of the MCMC corner plots are listed in Appendix Figure \ref{fig:mcmc_1124aa} -- \ref{fig:mcmc_sgr1935}. From the fitting results, we gain the following insight about FRBs.

\begin{table*}[!ht]
\centering
\hspace*{-3.7cm}
\addtolength{\tabcolsep}{-5pt}
\begin{tabular}{lllllllllllll}
    \hline
    Source/Parameter & $\alpha$ & $\beta$ & $\log P$ & $\Gamma$ & $\rho_{max}$ & $\log \mu$ & $\log K$ & $a$ & $\log c$ & $p$ & $\beta_e$ & $m_c$\\
    \hline
FRB 20121102A & $2.63^{+0.54}_{-1.07}$ & $2.43^{+0.62}_{-1.36}$ & $0.14^{+2.69}_{-2.45}$ & $101.13^{+0.52}_{-0.55}$ & $18.04^{+0.95}_{-0.62}$ & $-1.97^{+0.02}_{-0.02}$ & $-2.80^{+0.40}_{-0.30}$ & $-0.55^{+1.37}_{-1.35}$ & $-2.56^{+0.14}_{-0.26}$ & $1.81^{+0.18}_{-0.22}$ & $3.94^{+0.59}_{-0.75}$ & $40.30^{+0.06}_{-0.06}$ \\
FRB 20201124A(A) & $3.19^{+1.44}_{-0.90}$ & $5.08^{+2.04}_{-2.18}$ & $0.74^{+0.63}_{-0.81}$ & $101.63^{+2.04}_{-2.99}$ & $19.43^{+2.00}_{-1.15}$ & $-2.15^{+0.02}_{-0.02}$ & $-3.05^{+0.43}_{-0.35}$ & $-0.58^{+1.68}_{-1.18}$ & $-0.76^{+0.18}_{-0.18}$ & $4.62^{+1.10}_{-1.56}$ & $1.35^{+0.11}_{-0.19}$ & $39.40^{+0.09}_{-0.15}$ \\
FRB 20201124A(B) & $3.83^{+0.54}_{-1.59}$ & $2.47^{+0.62}_{-0.44}$ & $0.51^{+1.24}_{-1.67}$ & $99.80^{+2.13}_{-3.48}$ & $19.79^{+1.72}_{-0.62}$ & $-1.17^{+0.04}_{-0.04}$ & $-1.92^{+0.52}_{-0.37}$ & $-0.48^{+1.76}_{-0.73}$ & $-0.86^{+0.28}_{-0.17}$ & $3.14^{+0.57}_{-1.40}$ & $1.08^{+0.06}_{-0.07}$ & $39.24^{+0.06}_{-0.07}$ \\
FRB 20220912A & $3.36^{+0.74}_{-0.87}$ & $2.27^{+3.14}_{-2.43}$ & $0.54^{+1.23}_{-1.85}$ & $97.46^{+17.53}_{-6.44}$ & $17.05^{+9.73}_{-4.36}$ & $-1.41^{+0.03}_{-0.04}$ & $-2.39^{+0.57}_{-0.62}$ & $-0.07^{+1.62}_{-1.02}$ & $-0.48^{+0.14}_{-0.18}$ & $5.81^{+1.18}_{-1.90}$ & $1.40^{+0.12}_{-0.18}$ & $39.14^{+0.66}_{-0.27}$ \\
FRB 20190520B & $6.19^{+4.95}_{-10.59}$ & $15.76^{+15.67}_{-29.14}$ & $-0.02^{+2.13}_{-2.41}$ & $323.71^{+210.69}_{-340.84}$ & $24.00^{+16.37}_{-18.44}$ & $-2.24^{+0.14}_{-0.24}$ & $-3.16^{+0.92}_{-0.70}$ & $0.09^{+1.95}_{-1.78}$ & $-0.68^{+0.37}_{-0.17}$ & $5.77^{+1.94}_{-2.19}$ & $6.19^{+2.56}_{-2.20}$ & $42.40^{+2.03}_{-1.01}$ \\
SGR J1935+2154 & $38.35^{+8.26}_{-14.42}$ & $1.03^{+1.51}_{-1.36}$ & $0.51^{+0.03}_{-0.01}$ & $53.42^{+8.84}_{-16.29}$ & $0.71^{+0.36}_{-0.71}$ & $-0.21^{+0.12}_{-0.10}$ & $-3.52^{+1.15}_{-0.91}$ & $0.34^{+2.33}_{-1.70}$ & $-2.39^{+0.41}_{-0.32}$ & $2.23^{+0.69}_{-0.67}$ & $7.04^{+1.50}_{-1.90}$ & $26.93^{+0.17}_{-0.59}$ \\

\hline
\end{tabular}
\addtolength{\tabcolsep}{5pt}
\caption{List of best-fit and 16th and 84th percentile values for the parameters for different FRBs and SGR J1935+2154. $\log P$ stands for base 10 logarithm for the spin period.}
\label{table:fit_parameter}
\end{table*}

In general, FRBs have smaller $\alpha$ and larger $\rho_\mathrm{max}$ than the pulsed emission from SGR J1935+2154. In contrast, the pulsed emission from SGR J1935+2154 exhibits a large $\alpha$ and very small $\rho_\mathrm{max}$. Consequently, bursts from FRB sources can be observed regardless of the rotational phase, while the radio pulses from SGR J1935+2154 can only be observed within a small window. As a result, the periods of magnetar emitting FRBs cannot be easily detected, while the period of SGR J1935+2154 is significantly detected based on its pulsed emission. Assuming that the pulsed emission and FRB emission are directed at the same axis, the large inclination angle for SGR J1935+2154 could also explain the relatively lower observed energy of its FRBs \citep[e.g.][]{kirsten2021DetectionTwoBright}. Magnetars with higher $\alpha$ may have difficulties producing FRBs, or FRBs could not escape the magnetospheres of high $\alpha$ magnetars \citep[e.g.][]{qu2022TransparencyFastRadio}.

The best-fit value for the period of SGR J1935+2154 in this study is $\sim3.258^{+0.107}_{-0.201}\,\mathrm{s}$, which has a $\sim0.3\%$ error when compared with the reference value of $\SI{3.24781628}{\s}$ as measured from timing observations \citep{zhu2023RadioPulsarPhase}. The small inclination angle for FRBs was also recently pointed out by \cite{beniamini2025RoleMagneticRotation}.

Intriguingly, the best-fit values of $p$ are rather high, while $a$ is essentially zero. This is significantly different from earthquakes, whose $p$ typically takes values between $1$ and $1.5$, and $a$ is usually greater than $1$ \citep[e.g.][]{utsu1995CentenaryOmoriFormula,omi2014EstimatingETASModel,seif2017EstimatingETASEffects}. As pointed out by \cite{totani2023FastRadioBursts}, a higher $p$ means that the aftershock rate of FRBs decays faster than that of earthquakes after the mainshock. Thus, the waiting times between mainshocks and aftershocks are more separated in FRBs, making the bimodality in the waiting time distribution more significant.

On the other hand, $a$ represents the degree of correlation between the number of aftershocks and the magnitude of the mainshock, as defined in Equation \ref{eq:activation_function}. The best-fit values being close to zero mean that the number of aftershocks is irrelevant to the energy of the mainshock FRB. This matches the findings of \cite{zhang2024ArrivalTimeEnergy} that FRBs do not show clustering in energy space as opposed to earthquakes.

When comparing the two episodes of FRB 20201114A, we find that while the background rate $\mu$ and aftershock rate $K$ differ due to varying activity levels, their spin period, spectral index $\beta_e$, and aftershock parameters $c$ and $p$ remain consistent with each other. Since the aftershock parameters are possibly linked to the physical properties of the magnetar crust, this consistency implies that the magnetar crust did not undergo significant changes during the two episodes. 

Another notable result is that most parameters of FRB 20121102A are significantly different from other FRBs. This is consistent with the results reported by \cite{sang2024QuantifyingRandomnessScale}, which suggest that the behavior of FRB 20121102A in the time--energy space is unique compared to other FRBs. The characteristic time $c$ and aftershock decay index $p$ are smaller than those of other FRBs, while the spectral index $\beta_e$ is higher. This implies that the magnetar emitting FRB 20121102A may have different physical or environmental properties compared to other FRB-emitting magnetars.

Finally, while the numerical values of $\mu$ and $c$ parameters for earthquakes and FRBs occupy similar orders of magnitude when normalized by their respective time units \citep[e.g.,][]{utsu1995CentenaryOmoriFormula,omi2014EstimatingETASModel,seif2017EstimatingETASEffects}, their actual values differ by $\sim5$ orders of magnitude due to different units (days vs seconds). This significant distinction reveals that FRB-related processes (potentially magnetar starquakes) operate on drastically compressed timescales compared to tectonic processes on the Earth, corresponding to their extreme astrophysical environments.

We note that a very significant degeneracy is present in the model parameters. Therefore, the best-fit parameters we present might only be one of the many possible combinations. For instance, $\Gamma$, $m_c$ and $\beta_e$ all govern the energy spectrum. A higher $\Gamma$ paired with a higher $m_c$ could produce roughly the same waiting time and energy distributions compared with when both are lower, as the effects of changing the two parameters cancel out. Similar degeneracies are present in other groups like $\alpha$, $\beta$ and $\rho_{max}$ which are all related to the viewing angle distribution. These degeneracies arise because multiple parameter combinations can produce statistically indistinguishable fits to the observed waiting time and energy distributions. While this limits our ability to uniquely determine individual parameters, the bimodal waiting time distribution and the lower $\alpha$ leading to the non-detection of periodicity remain robust outcomes of the model. Future studies utilizing more types of data or refined analysis could help break these degeneracies.

Note that while parameters such as $\beta$, $\Gamma$, $\beta_e$ and $m_c$ for FRB 20190520B appear nominally distinct from those of other FRBs, their substantial uncertainties make them statistically consistent with the broader population. This apparent discrepancy likely stems from the relatively poor constraints on this source's energy distribution and its limited burst sample size rather than representing a genuine physical outlier.

\subsection{Waiting time and energy distributions}

\begin{figure*}
\centering
\subfloat[FRB 20201124A(A)]{
    \includegraphics[width=0.48\textwidth]{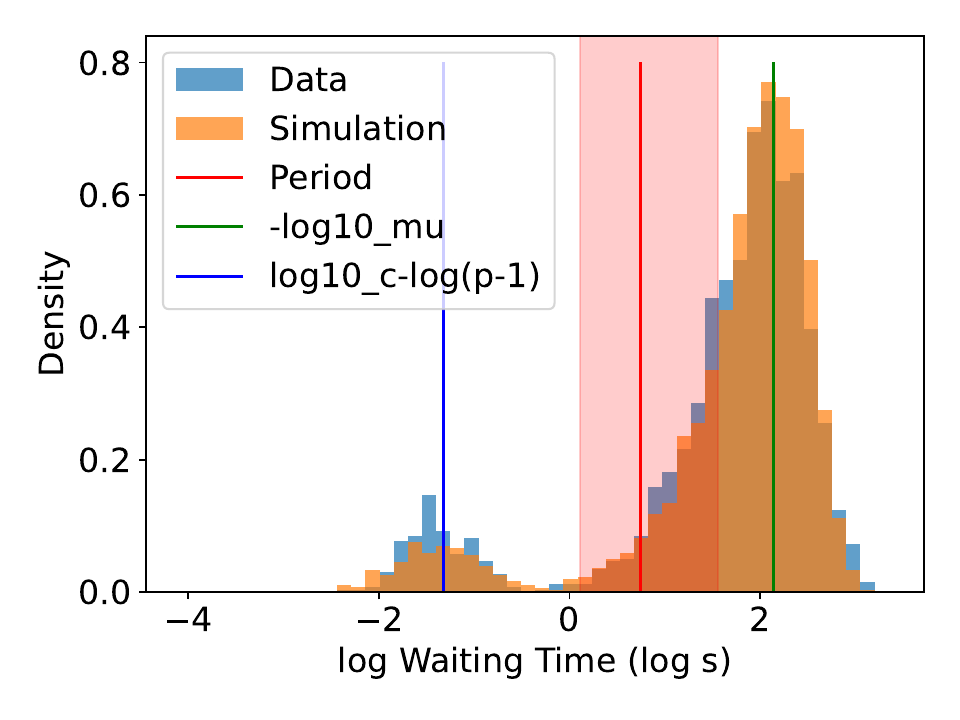}
    \label{fig:wt_1124aa}}
\subfloat[FRB 20201124A(B)]{
    \includegraphics[width=0.48\textwidth]{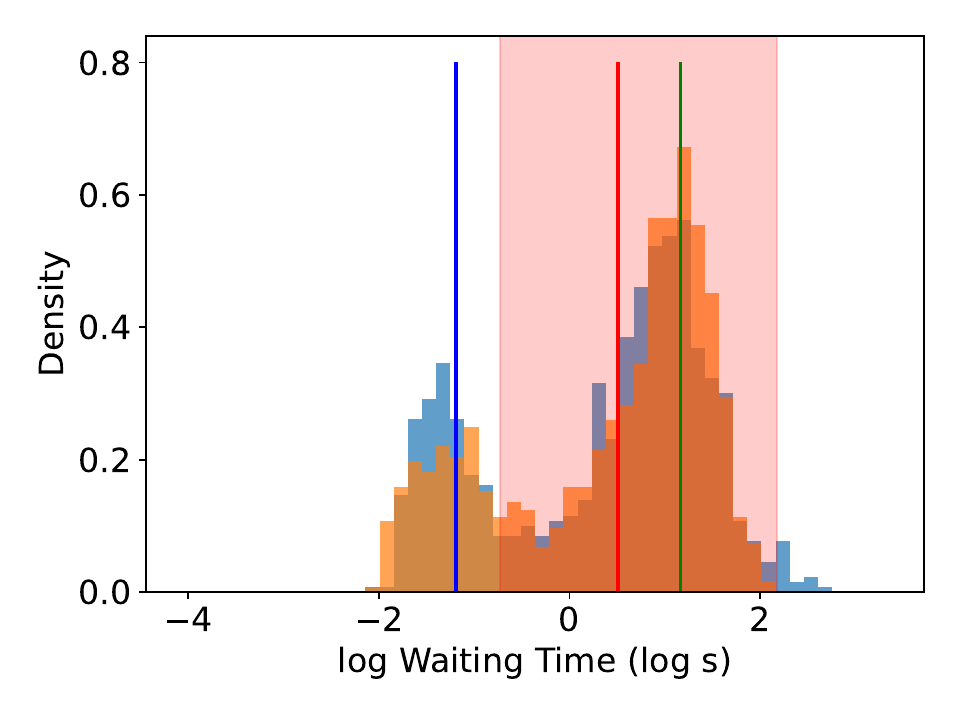}
    \label{fig:wt_1124ab}}\\
\subfloat[FRB 20121102A]{
    \includegraphics[width=0.48\textwidth]{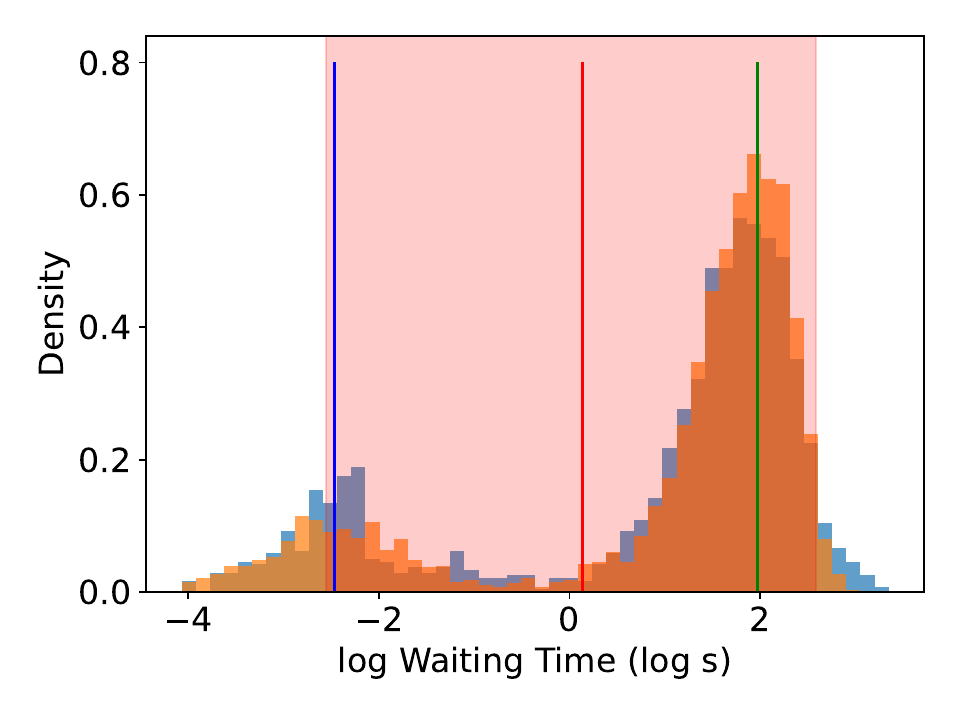}
    \label{fig:wt_121102}}
\subfloat[FRB 20190520B]{
    \includegraphics[width=0.48\textwidth]{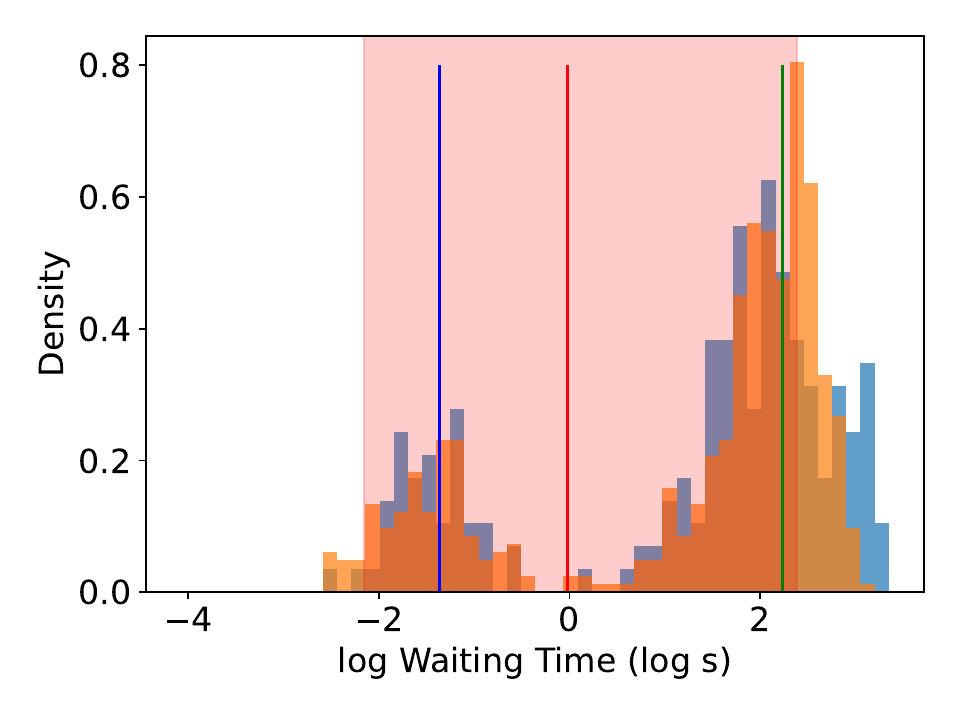}
    \label{fig:wt_190520}}\\
\subfloat[FRB 20220912A]{
    \includegraphics[width=0.48\textwidth]{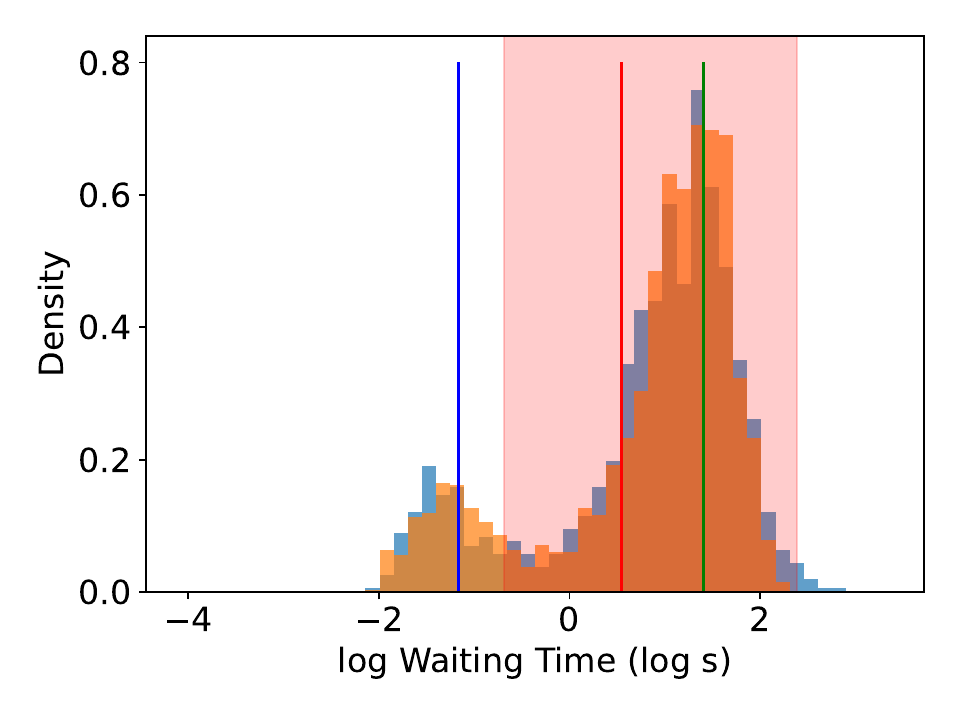}
    \label{fig:wt_220912}}
\subfloat[SGR J1935+2154]{
    \includegraphics[width=0.48\textwidth]{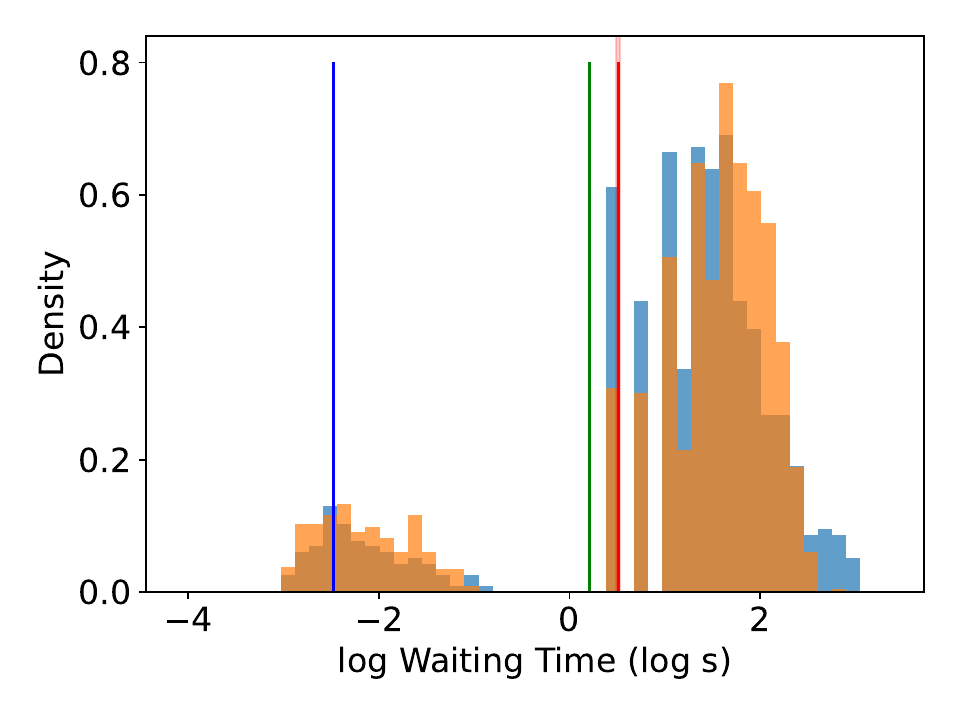}
    \label{fig:wt_sgr1935}}
\caption{Simulated and observed waiting time distributions of different FRBs and the pulsar phase of SGR J1935+2154. The best-fit spin periods are marked by red vertical lines, while the reciprocal of the background event rate $-\log{\mu}$ and $\log c-\log(p-1)$ are marked by green and blue lines respectively. The 16th and 84th percentile error ranges of spin periods are marked by red shaded areas. Note that the error region for the spin period of SGR J1935+2154 is too small to be shown.}
\label{fig:wt_distribution}
\end{figure*}

\begin{figure*}
\centering
\subfloat[FRB 20201124A(A)]{
    \includegraphics[width=0.48\textwidth]{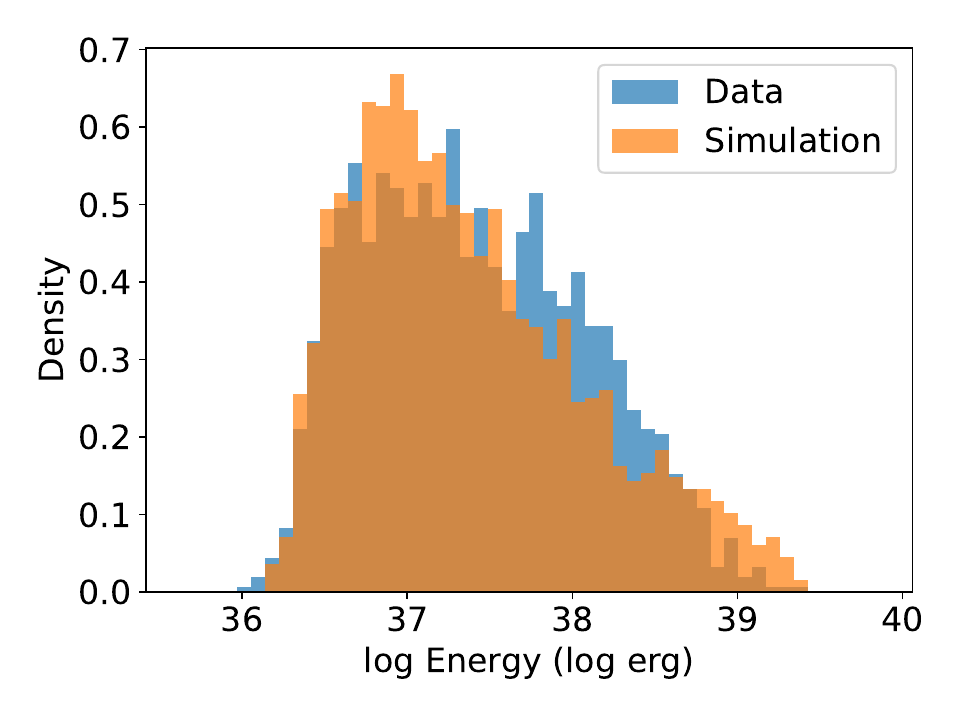}
    \label{fig:e_1124aa}}
\subfloat[FRB 20201124A(B)]{
    \includegraphics[width=0.48\textwidth]{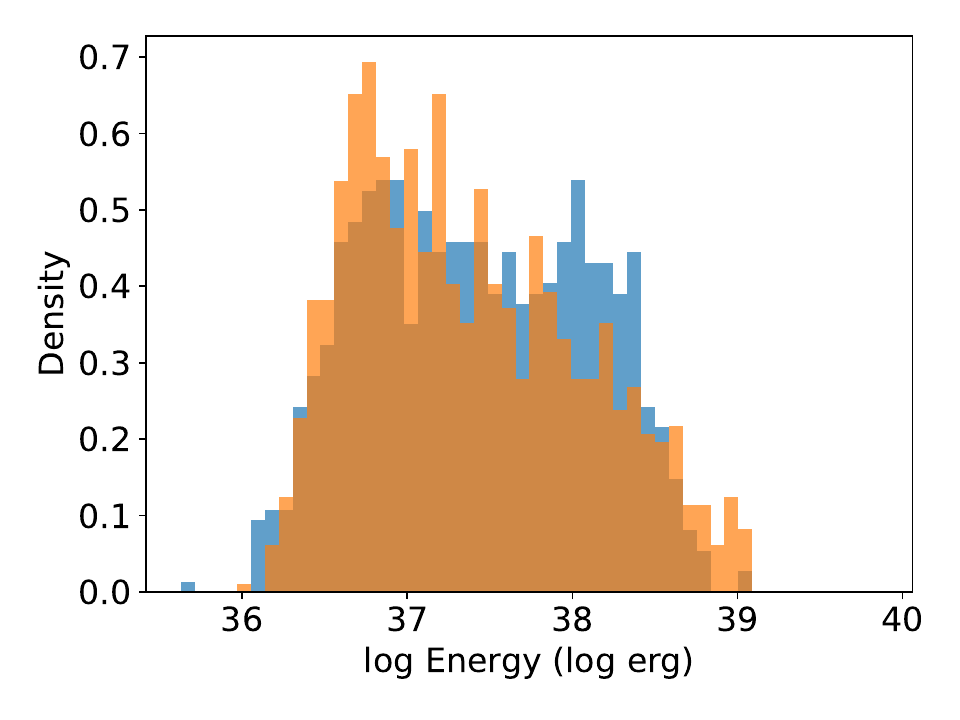}
    \label{fig:e_1124ab}}\\
\subfloat[FRB 20121102A]{
    \includegraphics[width=0.48\textwidth]{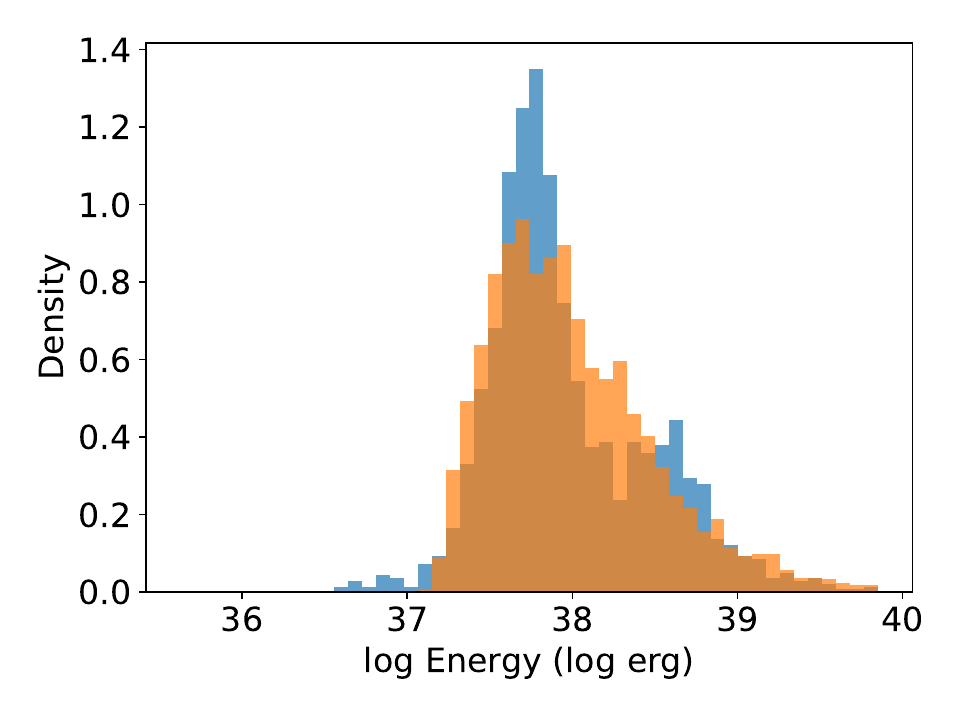}
    \label{fig:e_121102}}
\subfloat[FRB 20190520B]{
    \includegraphics[width=0.48\textwidth]{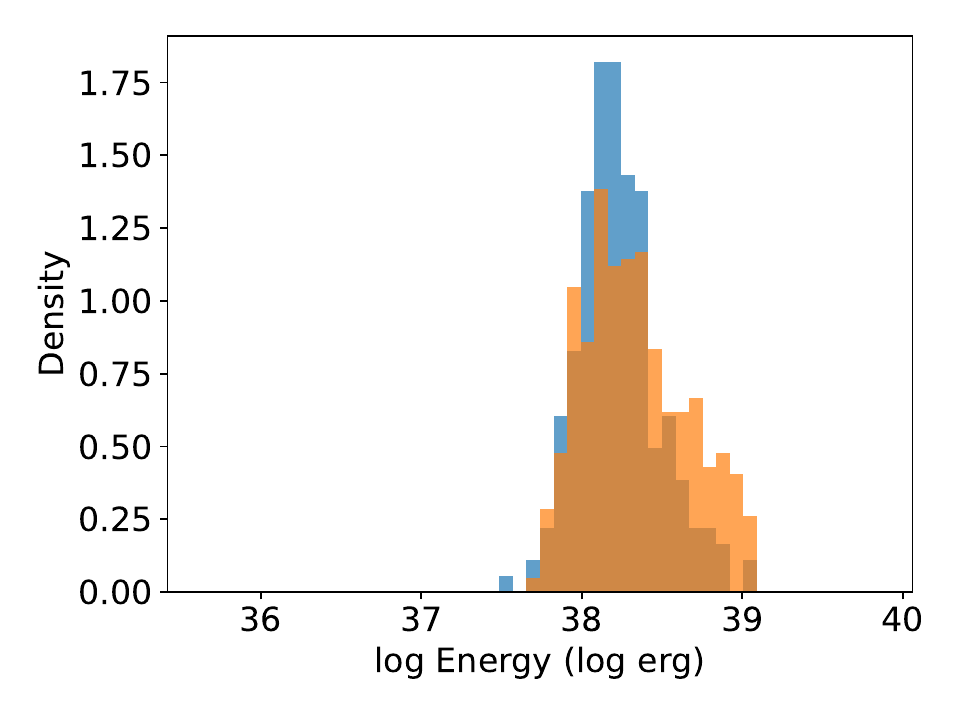}
    \label{fig:e_190520}}\\
\subfloat[FRB 20220912A]{
    \includegraphics[width=0.48\textwidth]{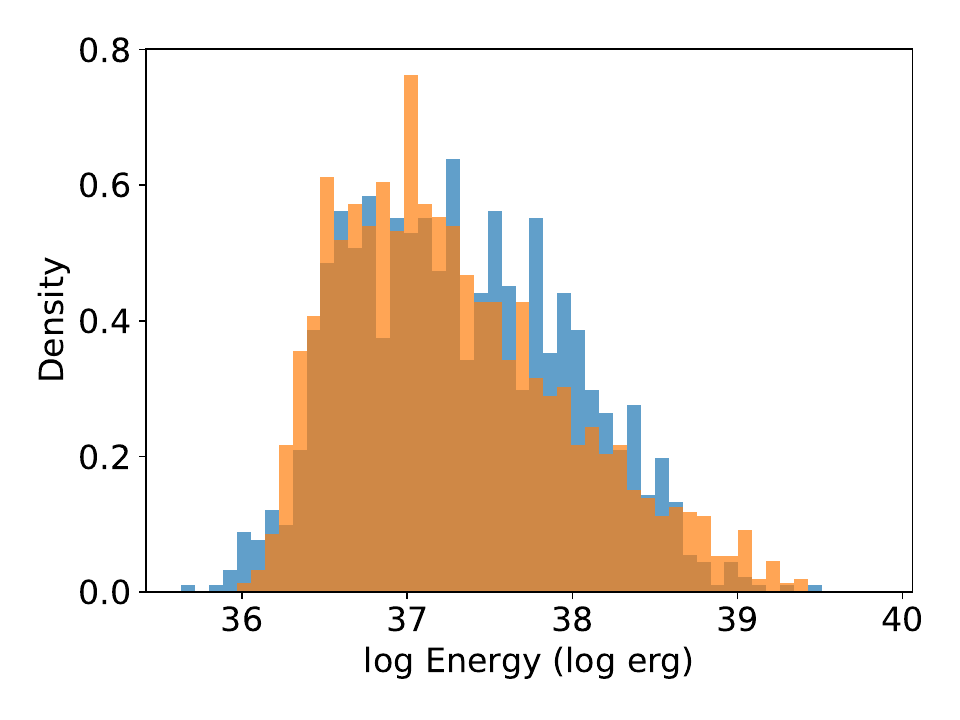}
    \label{fig:e_220912}}
\subfloat[SGR J1935+2154]{
    \includegraphics[width=0.48\textwidth]{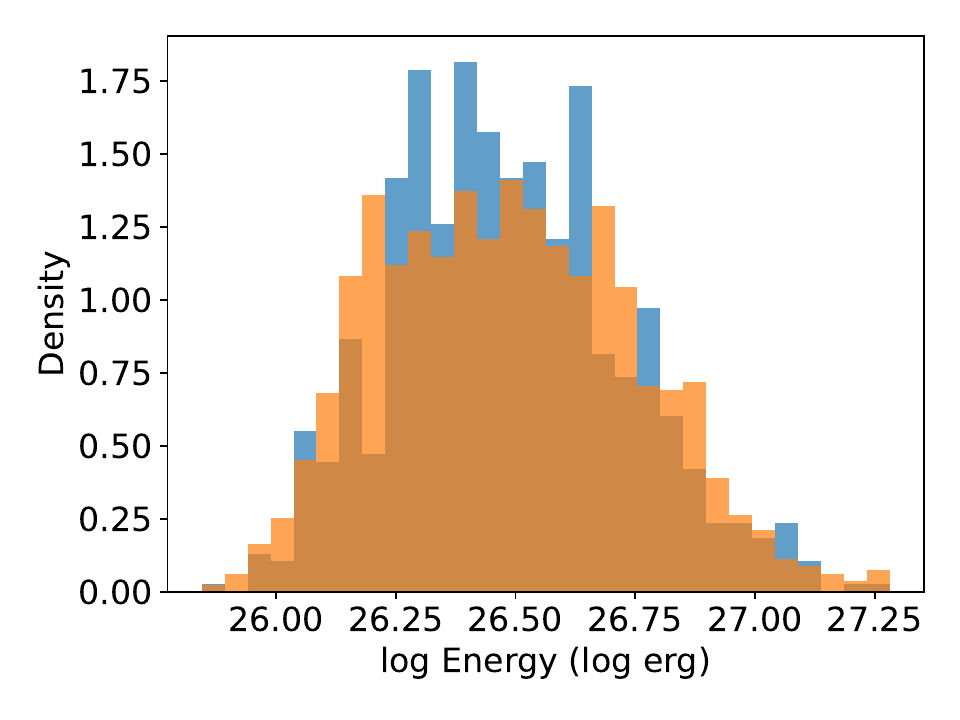}
    \label{fig:e_sgr1935}}
\caption{Simulated and observed energy distributions of different FRBs and the pulsar phase of SGR J1935+2154.}
\label{fig:e_distribution}
\end{figure*}

The bimodal distribution of FRB burst waiting times can be mostly well reproduced by our model. The simulated and observed waiting time and energy distributions are shown in Figures \ref{fig:wt_distribution} and \ref{fig:e_distribution}.

\begin{figure}
    \centering
    \includegraphics[width=0.49\textwidth]{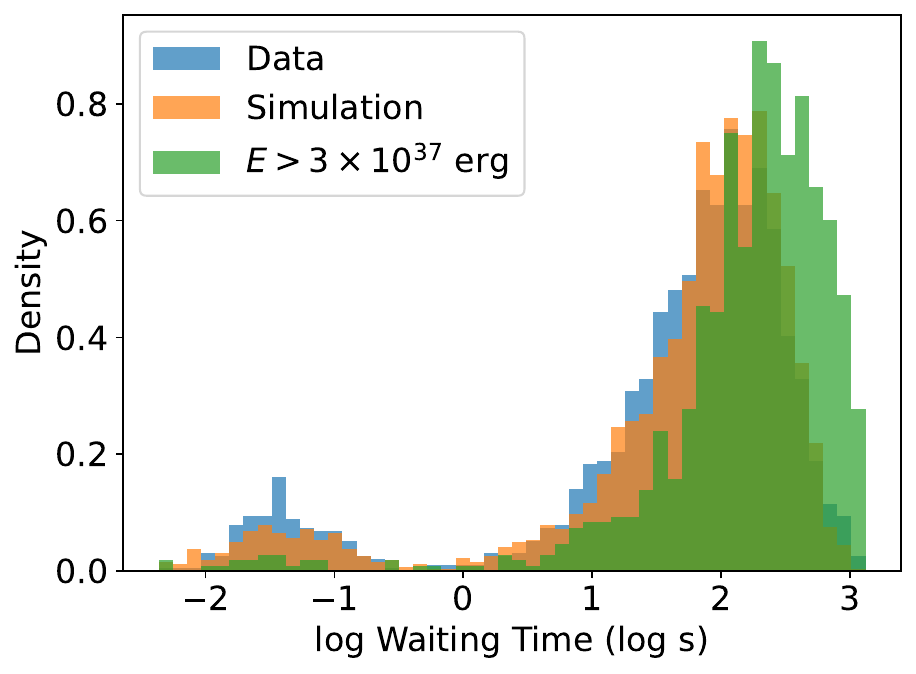}
    \caption{An example of all and high energy ($E>\SI{3e37}{\erg}$) simulated and observed waiting time distributions of FRB 20201124A(A).}
    \label{fig:wt_he}
\end{figure}

An interesting feature of FRB waiting times reported by \cite{li2021BimodalBurstEnergy} is that increasing the energy threshold only moves the right peak to larger values, but not the left peak. As shown in Figure \ref{fig:wt_he}, our model can reproduce this phenomenon. The reason behind this is that the right peak mainly corresponds to the background events (mainshocks), and a higher energy threshold significantly decreases the observed event rate. The left peak, on the other hand, mainly corresponds to triggered events (aftershocks). Although a higher energy threshold makes the triggering events of the aftershocks unobservable, their triggering effect still persists, thus the waiting times of the triggered events are less affected.

Similar phenomenon is reported by \cite{konijn2024NancayRadioTelescope} on observations of FRB 20220912A with the Nan\c{c}ay Radio Telescope, which has a lower sensitivity and thus a higher detection threshold compared to FAST. They found that the location of the left peak in the waiting time distribution is consistent with FAST observations, while the right peak moved to the right due to lower event rate.

As marked by green vertical lines in Figure \ref{fig:wt_distribution}, the locations of the right peaks in the waiting time distributions almost perfectly align with the value of $-\log{\mu}$. Since $\mu$ is the event rate for background events, $-\log{\mu}$ represents the logarithm of the average expected waiting time of a stationary Poisson process with an event rate of $\mu$. This shows that the right peak is dominated by background events. 

The locations of the left peaks as marked by blue vertical lines in the waiting time distributions for most FRBs are at about $\SI{1.5e-1}{\s}$, with the exception of FRB 20121102A. Additionally, the location of the left peak seems to align with $\log c-\log(p-1)$. This shows that the left peak is related to the aftershock parameters and potentially the physical parameters of the magnetar crust. A more detailed derivation of the shape of the distribution and the location of the peaks is in Section \ref{subsec:peak_location}. The fact that the location of the left peak does not rely on the event rate also supports the results shown in Fig. \ref{fig:wt_he}, where a higher energy threshold only moves the right peak but not the left one.

Our model predicts some waiting times at orders of $\SIrange{e-4}{e-5}{\s}$. Since the typical duration of the burst pulses of repeating FRBs is of the order of $\sim\SI{e-3}{\s}$, those spikes with shorter waiting times would not be discernible and we simply filter out simulated waiting times below the shortest observed waiting time.

\begin{figure}
    \centering
    \includegraphics[width=0.49\textwidth]{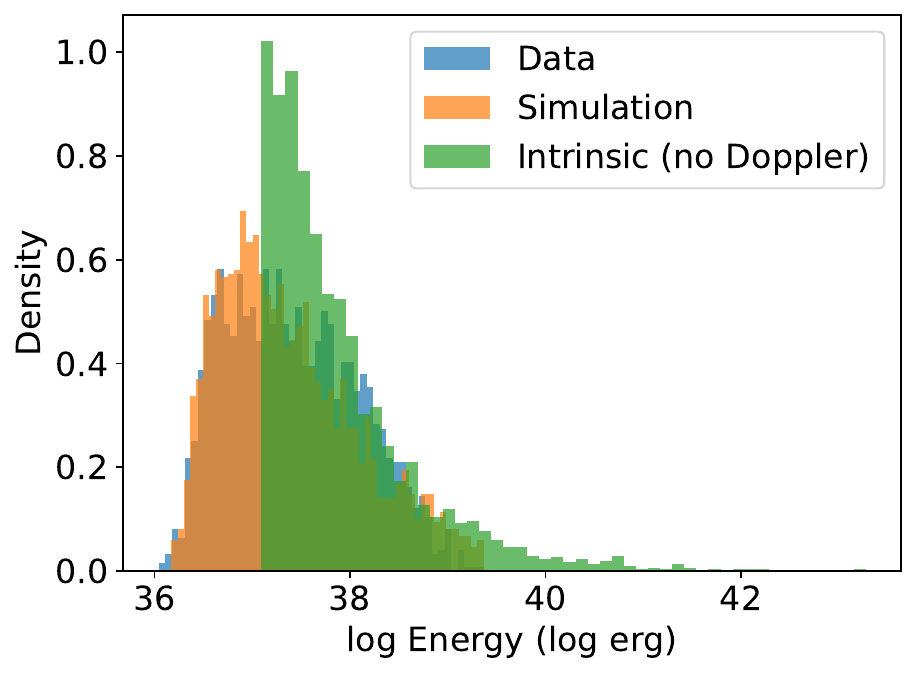}
    \caption{Comparison of observed, simulated observed and observed intrinsic energy (no Doppler effect) distributions of FRB 20201124A(A).}
    \label{fig:e_1124_intrinsic}
\end{figure}

The energy distribution is somewhat reproduced but not as well as the waiting time distribution, especially for FRB 20190520B. Notably, the bimodal distribution is not significantly reproduced. This could suggest that some factors other than the viewing angle effect are at play to generate the bimodal distribution of energy. In any case, it is interesting to note that we can get energy distributions similar to the observed ones simply from an intrinsic energy with power-law distribution but with the viewing angle effect included. A comparison among the observed, simulated observed, and observed intrinsic energy (without Doppler effect) distributions of FRB 20201124A(A) is shown in Figure \ref{fig:e_1124_intrinsic}.

\subsection{Location of peaks in the waiting time distribution}
\label{subsec:peak_location}

In general, rotation modulation does not have a large impact on the waiting time distribution for the hyper-active repeaters, but it does have a great impact on the energy distribution. Therefore the waiting time distribution can be well described by only the ETAS model. Furthermore, rotation modulation does have a large impact on the waiting time distribution of the pulsar phase of SGR J1935+2154. The fact that our model can reproduce very different waiting time distributions just by changing the parameters shows the strength of our model.

The right peak is dominated by background events, which follow a stationary Poisson process with a mean event rate of $\mu$. It is well known that the waiting time between events from stationary Poisson processes follows an exponential distribution. The probability density function (PDF) of the waiting time distribution of the background event is then:
\begin{equation}
    f_b(\tau)=\mu e^{-\mu\tau}.
\end{equation}

We are more interested in the distribution of the logarithm of the waiting time, which can be derived from the cumulative density function (CDF):
\begin{equation}
    F_b(\tau)=\int_0^\tau f_b(\tau') d\tau'= 1-e^{-\mu\tau}.
\end{equation}

Replace $\tau$ with its base 10 logarithm, $\mathcal{T}=\log \tau$, we can write the CDF for the logarithm of the waiting time of background events:
\begin{equation}
    F_b(\mathcal{T})=1-e^{-\mu\cdot10^{\mathcal{T}}}.
\end{equation}

The PDF can then be derived by taking the derivative of the CDF:
\begin{equation}
    f_b(\mathcal{T})=\dv{F_b(\mathcal{T})}{\mathcal{T}}=\mu\ln10\cdot10^{\mathcal{T}}e^{-\mu\cdot10^{\mathcal{T}}}.
\end{equation}

We can then obtain the location of the right peak by letting the derivative of the PDF $f_b(\mathcal{T})$ be 0. The right peak is at $\mathcal{T}=-\log \mu$.

Since the branching ratio $\eta$ is relatively small for the FRBs, most background events do not generate more than one aftershock and the number of higher generation aftershocks is small. Therefore, the left peak is dominated by the waiting time between background events and their first aftershocks. The PDF of the waiting times in the left peak can then be simply derived by normalizing the activation function $g(\Delta t,m)$ defined in Eq. \ref{eq:activation_function}:
\begin{equation}
    f_a(\tau)=\frac{g(\tau,m)}{\int_0^{\infty} g(\tau,m) d\tau}=\frac{g(\tau,m)}{n_{AS}(m)}=\frac{(p-1)c^{p-1}}{(\tau+c)^p}.
\end{equation}

The CDF is then:
\begin{equation}
    F_a(\tau)=\int_0^\tau f_a(\tau') d\tau'= 1-\qty(\frac{c}{c+\tau})^{p-1}.
\end{equation}

Again, introducing $\mathcal{T}=\log \tau$:
\begin{equation}
    F_a(\mathcal{T})= 1-\qty(\frac{c}{c+10^{\mathcal{T}}})^{p-1},
\end{equation}
and taking derivative, one obtains
\begin{equation}
    f_a(\mathcal{T})= \dv{F_a(\mathcal{T})}{\mathcal{T}}=\frac{(p-1)\ln 10}{c}\cdot 10^{\mathcal{T}}\cdot \qty(\frac{c}{10^{\mathcal{T}}+c})^p.
\end{equation}

The location of the right peak is at $\mathcal{T}=\log c-\log(p-1)$ by letting the derivative of the PDF $f_a(\mathcal{T})$ to be 0.

\begin{figure}
    \centering
    \includegraphics[width=0.49\textwidth]{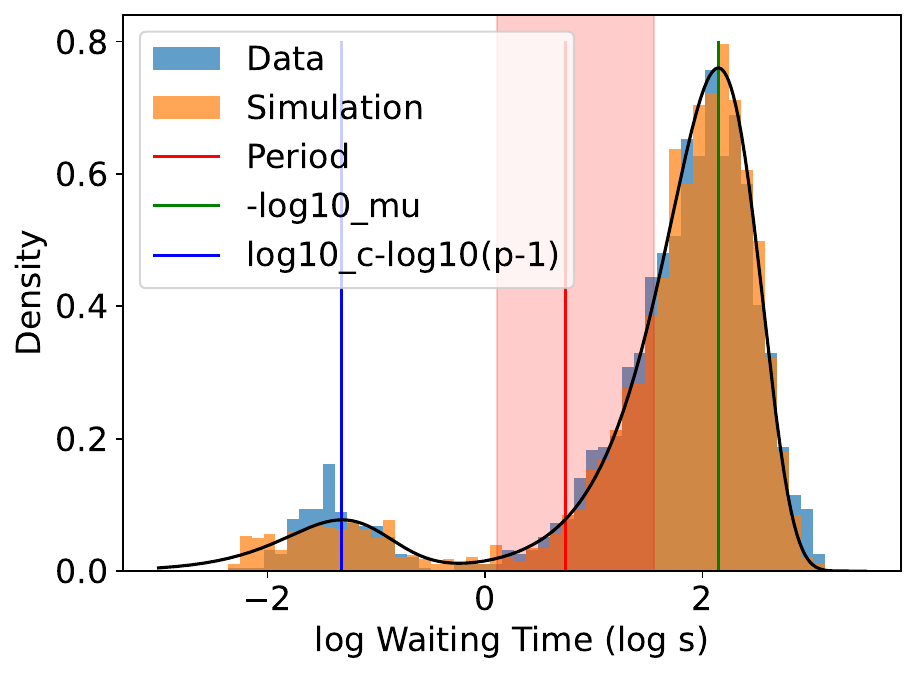}
    \caption{An example of the derived shape of the two peaks with fitted parameters of FRB 20201124A(A). The derived PDFs are shown by a black line. The left and right peak PDFs are normalized by the number of data points in the peaks accordingly. It can also be seen that the effect of rotation modulation on the waiting time is small, as is expected by the small $\alpha$ and large $\rho_{max}$ angles.}
    \label{fig:wt_1124a_fit}
\end{figure}

An example of the derived shape of the two peaks is shown in Fig. \ref{fig:wt_1124a_fit}.

\subsection{Periodicity and spin period}

As shown in Figure \ref{fig:wt_distribution}, while the spin periods cannot be very accurately determined, the best-fit spin periods and error regions for the FRBs generally lie between the two peaks in the waiting time distribution. Specifically, the best-fit spin periods are roughly between $\SIrange{1}{10}{\s}$ for all FRBs, inline with the spin periods of most magnetars \citep[e.g.][]{beniamini2019FormationRatesEvolution}.

The spin periods of the two episodes of FRB 20201124A are consistent with each other. Our model can find the period of SGR J1935+2154 with very low uncertainty and approximately $0.3\%$ error even without prior knowledge. This supports the robustness of our results for the periods.

\begin{figure}
    \centering
    \includegraphics[width=0.49\textwidth]{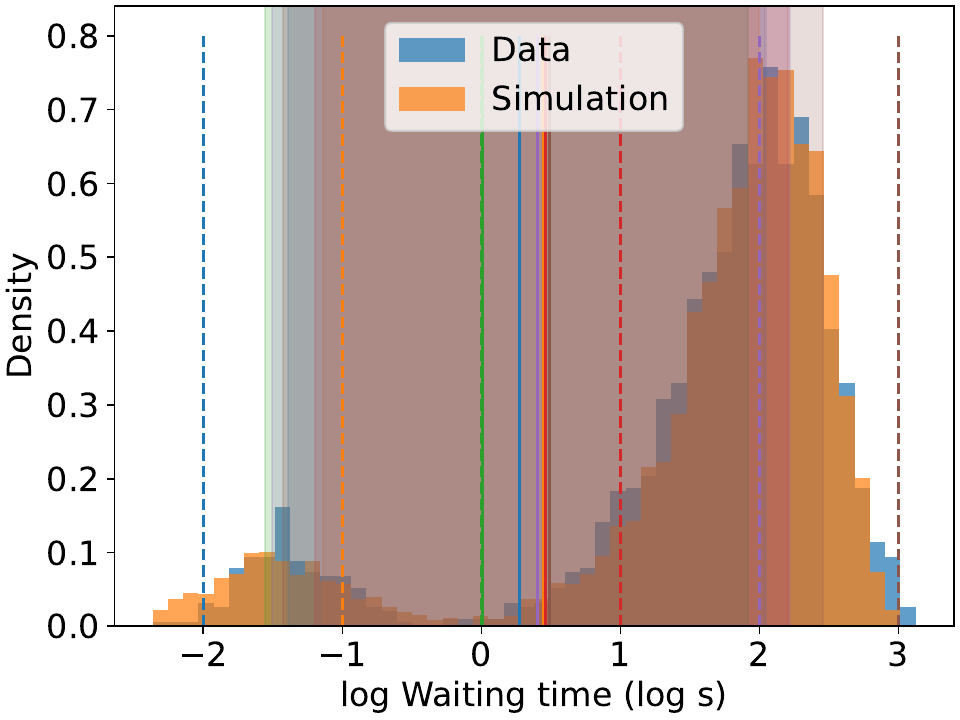}
    \caption{Fitting results for FRB 20201124A(A) with different initial spin period values. The dashed lines represent the initial values, the solid lines represent the best-fit values, and the shaded regions represent the 16th/84th percentile ranges. The lines and shades with the same color represent the groups with the same initial values.}
    \label{fig:1124a_periods}
\end{figure}

To show that the spin periods lying between the two peaks of the waiting time distribution are not a result of the initial values used in the MCMC fitting, we used different initial values for the spin period and conducted the same MCMC fitting for FRB 20201124A(A). The results are shown in Figure \ref{fig:1124a_periods}. The best-fit spin period values and corresponding error regions are roughly between the two peaks in the waiting time distribution despite different initial values, even for initial values away from the valley region.

For the spin period to locate in the valley region, it could be possible that the aftershocks happen close to the mainshocks on the magnetar, therefore the line of sight would miss the aftershocks if the spin period is too short. On the other hand, if the spin period is too long, the line of sight would not scan through enough active regions to generate the high event rate in hyper-active FRBs. Complex interactions involving the magnetosphere and the spin period could also play a role. A more in-depth analysis of the physical implications of the location of the spin period is left for future studies. Additionally, note that our model cannot test spin periods outside of the range of the observed waiting time.

\begin{figure}
    \centering
    \includegraphics[width=0.49\textwidth]{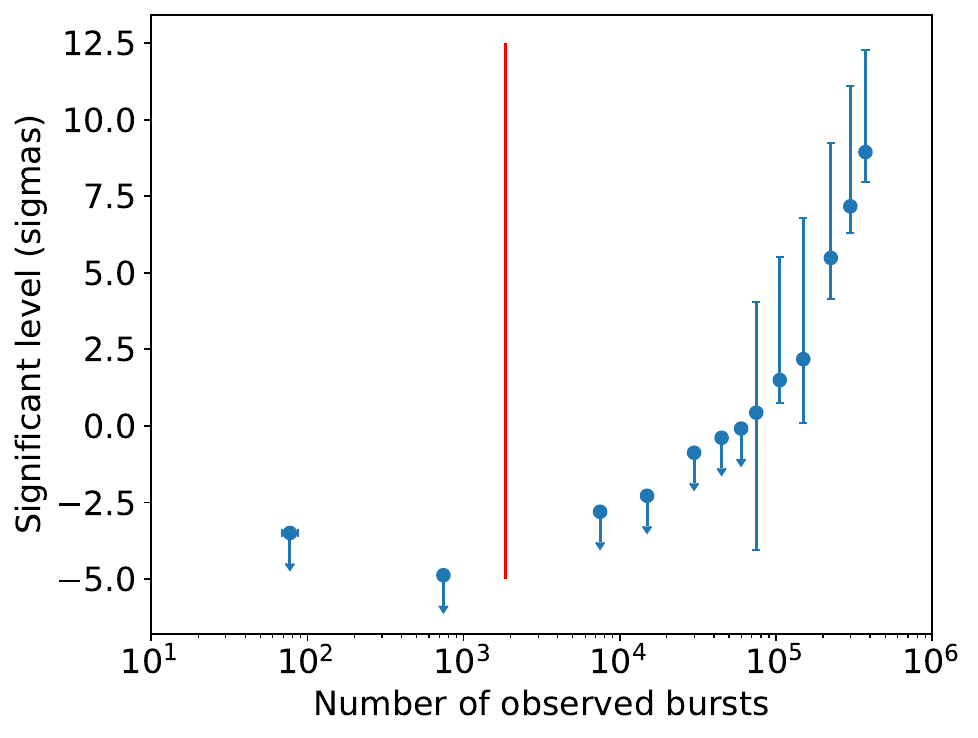}
    \caption{The relationship between the number of observed bursts and the level of significance of the period of FRB 20201124A(A). The number of bursts in the FRB 20201124A(A) dataset (1863) is shown with a red vertical line. Note that negative $\sigma$ results from the false alarm level being higher than 0.5.}
    \label{fig:1124aa_period_sinificance}
\end{figure}

Finally, we alter the total simulation time and consequently the total number of observable bursts in the simulations to test how many observed bursts are need to detect the period of FRBs at high confidence levels. We use the best-fit parameters of FRB 20201124A(A) and the false alarm level of the Lomb-Scargle periodogram to calculate the level of significance for the periodicity. The results are shown in Figure \ref{fig:1124aa_period_sinificance}.

More than $\num{e5}$ observed bursts are needed for the period to be detected at $3\sigma$ level, about $\num{2.9e5}$ observed bursts are needed for the period to be detected at $7\sigma$ level. The results from other FRBs are roughly the same, apart from FRB 20190520, whose parameters are poorly constrained. The reason why we chose FRB 20201124A(A) as an example in this section is that it has the most number of bursts and that its parameters are best constrained.

\section{Discussion}
\label{sec:discussion}

\subsection{To earthquake or not to earthquake}

There is an ongoing debate on whether the stochastic time-energy behaviors of FRBs resemble those of earthquakes. \cite{totani2023FastRadioBursts, wang2023RepeatingFastRadio,tsuzuki2024SimilarityEarthquakesAgain, gao2024ComparativeAnalysisScaleinvariant,yamasaki2025TimeFrequencyCorrelation} argued that FRBs are similar to earthquakes, while \cite{du2024ScalingUniversalityTemporal, zhang2024ArrivalTimeEnergy} claimed otherwise. One notable difference among these studies, apart from their conclusions, is that the studies asserting that FRBs resemble earthquakes primarily base their conclusions on the temporal behavior of FRBs, whereas those concluding that FRBs do not resemble earthquakes include energy in their analyses. 

In the model presented in this paper, the observed FRB energy is a function of the intrinsic energy and the viewing angle. Even though the observed energy distribution indeed does not satisfy the earthquake model prediction (consistent with \citealt{zhang2024ArrivalTimeEnergy}), because the viewing angle is stochastic, the observed FRB energy behavior is different from the intrinsic energy. Therefore, the intrinsic energy behavior could still resemble that of earthquakes. 

Furthermore, while this paper used the ETAS earthquake model to describe the dynamic of FRBs, many of the fitted ETAS parameters for FRBs are significantly different from those of earthquakes as pointed out in Section \ref{subsec:parameters}. Other factors could also distort the energy behaviors of FRBs since we did not perfectly reproduce the bimodal energy distribution of FRBs.

Additionally, \cite{sang2024QuantifyingRandomnessScale} conducted a similar analysis to \cite{zhang2024ArrivalTimeEnergy} but added Hurst exponent \citep{hurst1956PROBLEMLONGTERMSTORAGE} quantifying long-term memory in time series and non-Gaussian probability density distribution of fluctuations alongside Pincus index (also known as approximate entropy, \citealt{pincus1991ApproximateEntropyMeasure}) which assesses signal regularity, and found that long-term memory akin to earthquakes present in FRBs.

\subsection{Repeating and non-repeating FRBs}

\begin{figure*}
\centering
\subfloat[CHIME/FRB catalog 1]{
    \includegraphics[width=0.48\textwidth]{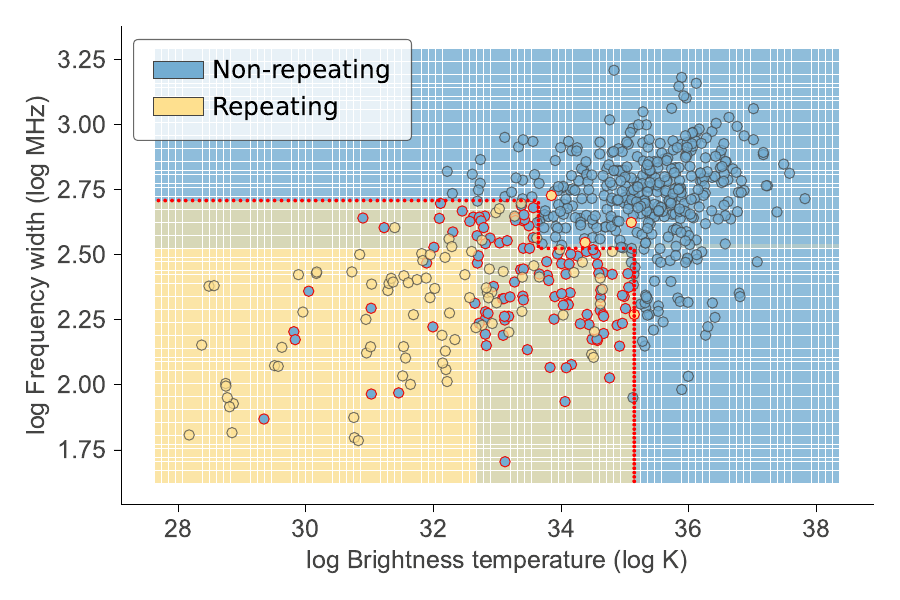}
    \label{fig:frb_ml_old}}
\subfloat[CHIME/FRB catalog 1 + CHIME/FRB 2023 catalog]{
    \includegraphics[width=0.48\textwidth]{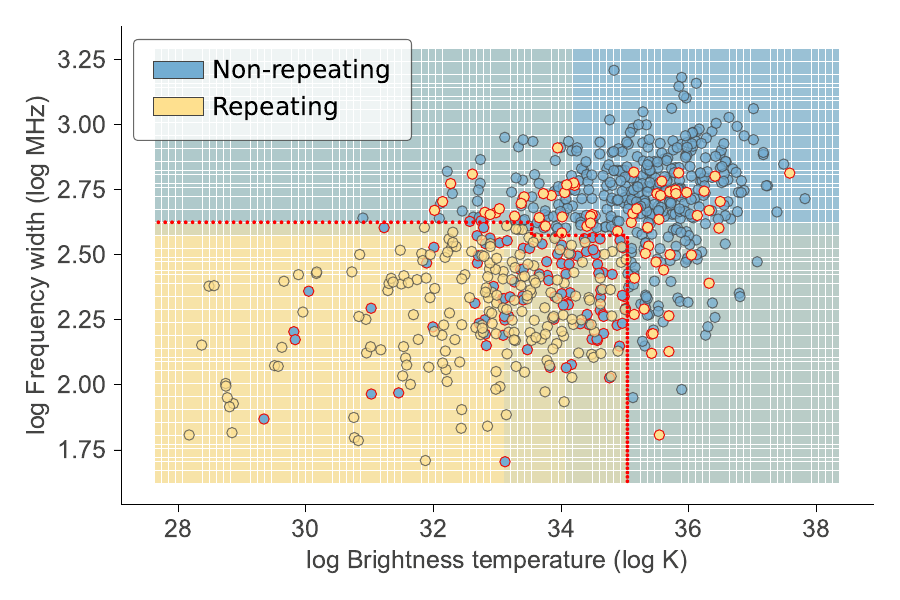}
    \label{fig:frb_ml_new}}
\caption{Comparison of the distribution of FRBs in the brightness temperature -- frequency bandwidth plane with or without FRBs from CHIME/FRB 2023 catalog. The points with red outlines are the ones that are misclassified by the machine learning algorithm. The red dashed line shows the decision boundary of the decision tree. The depth of colors represents the probability of either class. Many newly discovered repeating FRBs have high brightness temperatures and wide frequency bandwidth, and thus are more likely to be misclassified by the simple brightness temperature -- frequency bandwidth method.}
\label{fig:frb_ml}
\end{figure*}

Machine learning models and other analyses revealed that repeating FRBs in CHIME/FRB catalog 1 \citep{chime/frbcollaboration2021FirstCHIMEFRB} generally have lower brightness temperatures and narrower frequency bandwidths compared with non-repeaters \citep[e.g.][]{pleunis2021FastRadioBurst,luo2023MachineLearningClassification,zhu-ge2023MachineLearningClassification}. However, some new repeaters released by CHIME later showed higher brightness temperatures than those in CHIME/FRB catalog 1 \citep{thechime/frbcollaboration2023CHIMEFRBDiscovery}. A comparison of the distribution of FRBs in the brightness temperature -- frequency bandwidth plane with or without FRBs from CHIME/FRB 2023 repeater catalog is shown in Figure \ref{fig:frb_ml}.

According to our simulations, there are two possibilities to understand the results. First, higher brightness temperatures might correspond to higher intrinsic energies and lower intrinsic event rates, but the impact angle $\beta$ of those new repeaters is still low. Alternatively, wider frequency bandwidths may correspond to higher values of $\alpha$ and $\beta$, thus requiring the brightness temperatures of the FRBs to be high in order to be observed. In either case, it seems that the repeaters with higher brightness temperatures and wider frequency bandwidths take more time to be discovered.

In any case, while repeaters and non-repeaters still exhibit different distributions in brightness temperature, there is a significant overlap. As a result, the distinction between repeaters and non-repeaters is blurred. It could be that all FRBs are repeaters, but with different event rates and $\alpha$ and $\beta$ parameters, making the repetition of some harder to observe than some others.\footnote{\cite{liu2024GeometricNeutronStar} proposed a unified geometric model of FRBs that can possibly explain the difference in distributions of observational parameters of repeating and non-repeating FRBs. Their model is similar to the model presented in this paper, but their model predicts higher viewing angles for repeaters, opposite of what was proposed by this paper.}

\cite{wang2024MemoryBurstOccurrence} pointed out that there is a memory effect in the waiting time of FRBs in the sense that longer waiting times are often followed by longer waiting times and vice versa. This implies that the observed event rates of FRBs are tied to intrinsic parameters of FRB sources, such as the impact parameter $\beta$.

\begin{figure*}
\centering
\subfloat[Repeaters]{
    \includegraphics[width=0.48\textwidth]{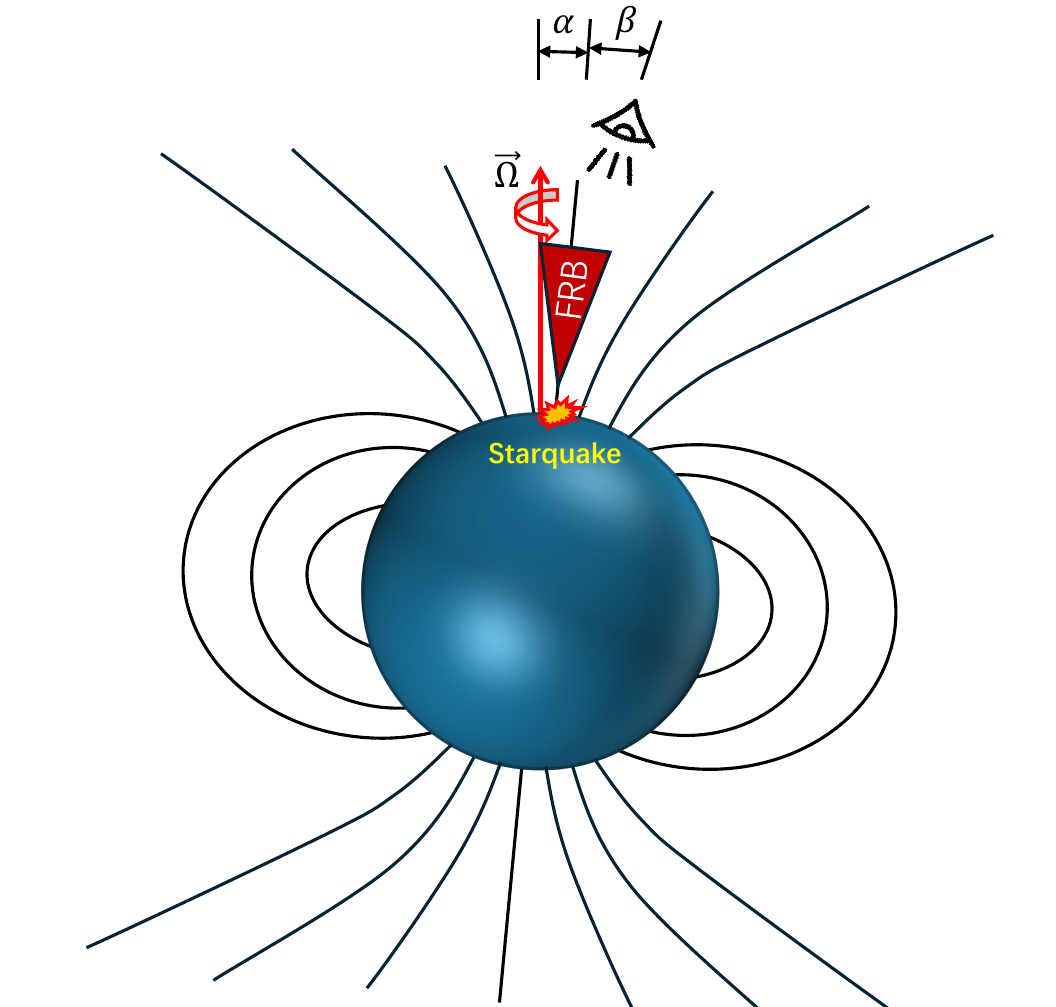}}
\subfloat[Non-repeaters]{
    \includegraphics[width=0.48\textwidth]{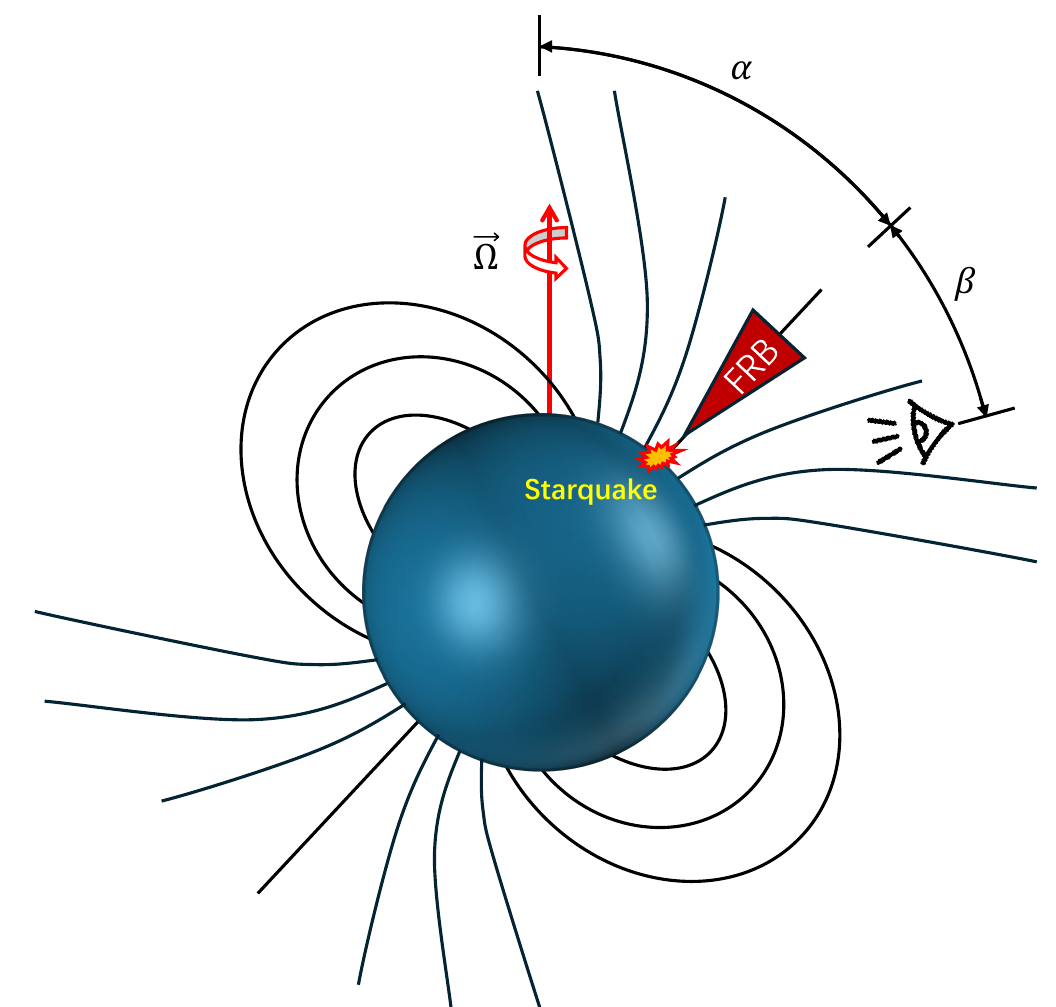}}
\caption{A cartoon sketch demonstrating the possible difference between repeaters and non-repeaters in the model presented in this paper. Repeaters shown on the left have low $\alpha$ and $\beta$, therefore more events can be observed, but the spin period is hidden. Non-repeaters shown on the right have high $\alpha$ and $\beta$, and very few events can be observed due to high viewing angles.}
\label{fig:cartoon}
\end{figure*}

If we assume that repeaters and non-repeaters all originate from magnetars with similar mechanisms, we can come up with a unified picture to understand their apparent differences. As shown in Figure \ref{fig:cartoon}, the main difference between repeaters and non-repeaters in our model lies on the different inclination angles: while the magnetic axes of repeaters are nearly aligned with the spin axis, those of apparent non-repeaters are likely more inclined. Note that both $\alpha$ and $\beta$ can cause difference in observed event rates. More detailed analysis to distinguish the effect of the two parameters is left for future studies. We note that \cite{beniamini2025RoleMagneticRotation} proposed a similar idea independently. 

Under the assumption that all FRBs are repeaters, we conduct a simulation with semi-global parameters to reproduce the repeater/non-repeater ratio in CHIME/FRB catalog 1 by altering the orientation of FRB-emitting magnetars.

We use all the fitted parameter of FRB 20201124A(A) and vary the impact parameter $\beta$. The simulation time is the total time for CHIME/FRB catalog 1, 342 days \citep{chime/frbcollaboration2021FirstCHIMEFRB}. The simulation generated 220322 FAST observable bursts. We consider an active window of 10\%, the total number of FAST observable bursts is $\sim22032$. Consider that the FRB 20201124A(A) catalog contains 1863 bursts observed in 82 hours and the active window, this corresponds to $\sim18648$ bursts in the same time range. The two numbers of bursts being comparable justifies our estimation of 10\% active window.

We then generate 10000 simulated FRBs. RA and DEC are generated isotropically, redshifts are generated assuming FRB sources have the same density in the comoving volume. We use a redshift range of \SIrange{0}{5.5}, as energy $<\SI{e42}{\erg}$ is not detectable by CHIME at this distance. DEC are generated in the range of \SIrange{-10}{90}{\degree} corresponding to the CHIME field of view. We use the same $\alpha$ as FRB 20201124A(A) for all the simulated FRBs, while $\beta$ are generated isotropically.

To convert the simulated energy to fluence, we use
\begin{equation}
    F=\frac{E(1+z)}{4\pi D_{\rm L}^2 \Delta\nu},
\end{equation}
where we adopted the CHIME bandwidth \SI{400}{\mega\hertz} as $\Delta\nu$.

We use a hard detection threshold of \SI{5}{\jansky\milli\s} \citep{chime/frbcollaboration2021FirstCHIMEFRB} for CHIME. We also consider the CHIME observation cadence provided in the CHIME/FRB catalog 1. For this simple simulation, we do not account for propagation effects and other observational biases related to burst duration, frequency bandwidth and others. Finally, we draw the number of observed bursts of each FRB source from Poisson distributions with means being their expected observable bursts.

From 10000 simulated FRB sources, we obtain 484 apparently non-repeating FRBs and 224 bursts from 79 repeating FRBs. Repeating source is $\sim14.03\%$ of all sources. In comparison, CHIME/FRB catalog 1 contains 474 non-repeating bursts, and 62 bursts from 18 repeating sources. Repeating source is $\sim3.66\%$ of all sources. Although the simulated repeater fraction is $\sim4\times$ the observed one, the two numbers are roughly consistent with each other, considering that most repeating FRBs are probably much less active and have higher $\alpha$ than FRB 20201124A. Additionally, some apparently non-repeaters with sub-bursts in the CHIME catalog might actually be repeaters in the model presented by this paper, thus the true repeater fraction might be higher in CHIME/FRB catalog 1. Due to the current lack of knowledge regarding the distributions of the FRB model parameters in the Universe, more detailed simulations are left for future investigations.

Nonetheless, purely from the point of view of the observed event rates, our model is compatible with the assumption that all FRB sources are repeaters with different parameters. It remains to be seen whether this assumption will stand true as we gain more insight into the distribution of the parameters of our model among FRB sources and the evolution of the fraction of repeating FRBs \citep{ai2021TrueFractionsRepeating}. It is also unknown that if our model can explain differences in other observed parameters like energy or frequency between repeaters and non-repeaters, as we mainly only considered event rate in this simulation. \citet{beniamini2025RoleMagneticRotation} gave some explanations to the parameter differences from the effects of $\alpha$ and $\beta$. More detailed analysis is left for future studies.

\subsection{Energy budget}
\citet{li2021BimodalBurstEnergy} estimated that the isotropic energy of 1652 bursts detected by FAST from FRB 20121102A in 2019 accounts for approximately 37.6\% of the total energy stored in the magnetosphere of the magnetar. Given that FRB 20121102A has been active for over a decade, the total isotropic energy of all the FRBs emitted by FRB 20121102A would be many times the available magnetic energy. Furthermore, bursts from a single day during an extremely active period of FRB 20201124A correspond to about 14.3\% of the magnetic energy in the magnetar's magnetosphere \citep{zhang2022FASTObservationsExtremely}. This severely challenges the magnetar model for hyper-active FRBs.

\citet{wang2024EnergyBudgetStarquakeinduced} proposed that other energy sources such as rotational energy or gravitational energy could be stored in the crust of a magnetar and be released in starquakes. These sources can provide energies much larger than the magnetic energy stored in the magnetosphere. In conjunction with our model, starquakes could trigger and power the particles that emit FRBs in the magnetosphere.

Another important factor to consider in calculating the total energy budget of FRBs is the beaming factor. In our model, all FRBs are beamed emission that roughly point to one direction. Therefore, the total energy of FRBs would be smaller than the isotropic energy calculated with the central energy of each FRB beam and the global beaming factor is smaller than 1. However, if the FRB is viewed off-axis, the observed energy would be smaller than the intrinsic energy. Consequently, the global beaming factor would be determined by the orientation of magnetic and spin axes and parameters like $\Gamma$ and $\rho_{\mathrm{max}}$.

On the other hand, while most FRBs from hyper-active repeaters can be observed, many of the apparently less active repeaters could have emitted a lot more FRBs that are not observed as the magnetic axes of their magnetars are not aligned with the line of slight. As a result, even for repeaters with few observed bursts, the total energy of their FRBs could still surpass the total energy budget of the magnetosphere and additional energy sources such as starquakes are required.

\section{Conclusions}
\label{sec:conclusions}

To address the critical challenges in magnetar models for FRBs, specifically the non-detection of spin periodicity and the observed bimodal waiting time distributions, we have developed a framework combining the ETAS earthquake model with RVM geometric constraints. Our methodology introduces rotational modulation into aftershock sequences, providing physical interpretations for both the temporal and energy distributions of FRBs while maintaining consistency with non-detection of rotational periodicity. Through a comprehensive analysis of multiple hyper-active repeating FRB sources and a pulsar phase from an FRB-emitting galactic magnetar, we reach the following key conclusions:

\begin{enumerate}
\item
    We can reproduce the bimodal waiting time distribution observed in hyper-active repeating FRB sources with the model presented in this paper. The left peak of waiting time corresponds to the parameters of the aftershock activation function, while the right peak corresponds to the background event rate.
\item
    Our model can also generate FRB energy distributions somewhat similar to the observed ones from a purely power-law distribution and the RVM model.
\item
    The best-fit value for the time offset parameter $c$ stays roughly constant in the two episodes of FRB 20201124A despite vastly different event rates, hinting that the characteristic time $c$ might be related to the physical properties of the magnetar crust. \cite{konijn2024NancayRadioTelescope} observed 696 Bursts from FRB 20220912A and found that the location of the left peak in the waiting time distribution is consistent with FAST observations, while the right peak moved to the right due to lower event rate. This is consistent with the location of the left peak being determined by parameters $c$ and $p$ that are related to the magnetar crust and stay roughly unchanged despite changes in background event rate $\mu$.
\item
    The period of FRB-emitting magnetars can be roughly constrained to be between the two peaks of the waiting time distribution. Moreover, the two episodes of FRB 20201124A yielded similar best-fit spin periods. Our model can also find the period of SGR J1935+2154 with very low uncertainty and approximately $0.3\%$ error without prior knowledge.
\item
    The magnetic inclination angle $\alpha$ must be small for non-detection of periodicity. Because the beaming angles of FRBs are very small, $\beta$ also need to be small for FRBs to be observed. Generally speaking, FRB-emitting magnetars are aligned rotators. A lot more bursts are needed for the period of FRB-emitting magnetars to be significantly detected. For FRB 20201124A(A) and the best-fit parameters, more than $\num{e5}$ observed bursts are needed for the period to be detected at the $3\sigma$ level. The conclusion regarding the small $\alpha$ was also recently and independently drawn by \citet{beniamini2025RoleMagneticRotation}. We further support this argument with a large sample of bursts detected from several distinct sources.
\item
    The magnetic inclination angle $\alpha$ for SGR J1935+2154 is larger than those of the hyper-active repeaters. Our model can reproduce the waiting time distribution of a pulsar phase of the magnetar that is very different from the hyper-active repeaters. This could explain the significant detection of the spin period and the relatively low observed energy for FRBs from the magnetar.
\item
    The magnitude sensitivity parameter $a$ is almost zero, meaning that the number of aftershocks is basically not correlated to the magnitude of the mainshock, a significant deviation from earthquakes on the Earth.
\item
     Our model is compatible with all FRB sources being repeaters with different parameters like $\alpha$, $\beta$ or $\mu$ resulting in different observed event rates. But whether this holds true remains to be seen as we learn more about parameter distributions and the evolving fraction of repeaters.
\end{enumerate}

\begin{acknowledgments}
    The authors thank Zhenghao Cheng, Ping Wang, Fayin Wang, Dongzi Li, Jinjun Geng, Bing Li, Renxin Xu, Yujia Wei, Qin Wu, Zenan Liu, Chenran Hu, Xiaohui Liu, Fen Lv, Chuanjie Zheng, Yuanhong Qu, Christopher Thompson, Leila Mizrahi, Tomonori Totani, Kiernan Folz-Donahue, Nicholas Gannon and Shaul Hurwitz for helpful discussions. JW Luo is supported by National Natural Science Foundation of China (grant no. 12403100), Hebei Natural Science Foundation (grant no. A2024205004), Hebei Yan-Zhao Golden Terrace Key Talent Program (Platform for Returnees from Overseas Studies, grant no. B2024017), Science Research Project of Hebei Education Department (grant no. QN2024287) and Science Foundation of Hebei Normal University (grant no. L2024B07).
\end{acknowledgments}

\bibliographystyle{aasjournal}
\bibliography{frb-waiting-time}

\appendix
\section{Glossary of symbols}
The following Table lists the symbols used in this paper and their descriptions.
\begin{table}[ht]
\centering
\begin{tabular}{ll}
    \hline
    Sign & Description \\
    \hline
    $E$ & Isotropic energy \\
    $F$ & Fluence \\
    $\Delta\nu$ & Frequency bandwidth \\
    $\tau_i$ & Waiting time between the $i$th and $i-1$th bursts \\
    $t_i$ & Time of the $i$th burst\\
    $\lambda$ & Earthquake event rate \\
    $m$ & Magnitude of earthquake, $m$=$\log E$ \\
    $f(m)$ & Probability distribution of earthquake magnitudes \\
    $\mu$ & Background event (mainshock) rate \\
    $g(\Delta t_i, m_i)$ & Aftershock activation function \\
    $\Delta t_i$ & Time difference between current time and earthquake $i$, $\Delta t_i=t-t_i$\\
    $m_c$ & Reference magnitude and the minimum magnitude \\
    $\beta_e$ & Power-law index for the Gutenberg–Richter law for magnitudes \\
    $K$ & Aftershock productivity parameter \\
    $a$ & Magnitude sensitivity parameter \\
    $c$ & Time offset parameter \\
    $p$ & Time decay exponent (Omori exponent) \\
    $n_{AS}$ & Number of expected aftershocks \\
    $\eta$ & Branching ratio \\
    $\alpha$ & Magnetic inclination angle \\
    $\beta$ & Impact parameter \\
    $\zeta$ & Line of sight colatitude, $\zeta=\alpha+\beta$ \\
    $\phi$ & Spin longitude\\
    $\rho_{max}$ & Maximum angle the FRB beam is allowed to deviate from the magnetic axis\\
    $\rho$ & Angular distance between the FRB beam and the magnetic axis\\
    $\theta_\rho$ & FRB beam relative longitude\\
    $\theta_{v0}$ & Viewing angle with $\rho_{max}=0$\\
    $\theta_{v}$ & Viewing angle considering $\rho_{max}$\\
    $P$ & Spin period\\
    $\mathcal{D}$ & Doppler factor\\
    $\mathcal{D}_0$ & Doppler factor at $\theta_v=0$\\
    $E_i$ & Intrinsic isotropic energy\\
    $E_o$ & Observed isotropic energy\\
    $\Gamma$ & Lorentz factor\\
    $\beta_d$ & Dimensionless speed\\
    $\mathcal{L}$ & Likelihood function for MCMC\\
    % $\mathcal{L}_{wt}$ & Likelihood for waiting time distribution\\
    % $\mathcal{L}_{E}$ & Likelihood for energy distribution\\
    % $\mathcal{L}_{p}$ & Likelihood for non-detection of periodicity\\
    \hline
\end{tabular}
\caption{Glossary of symbols used in this paper.}
\label{table:glossary}
\end{table}

\section{Detection probabilities}
The following Figures lists the detection probabilities and energy distributions of different FRBs and the pulsar phase of SGR J1935+2154 used in this paper.
\begin{figure*}[ht]
\centering
\subfloat[FRB 20201124A(A)]{
    \includegraphics[width=0.48\textwidth]{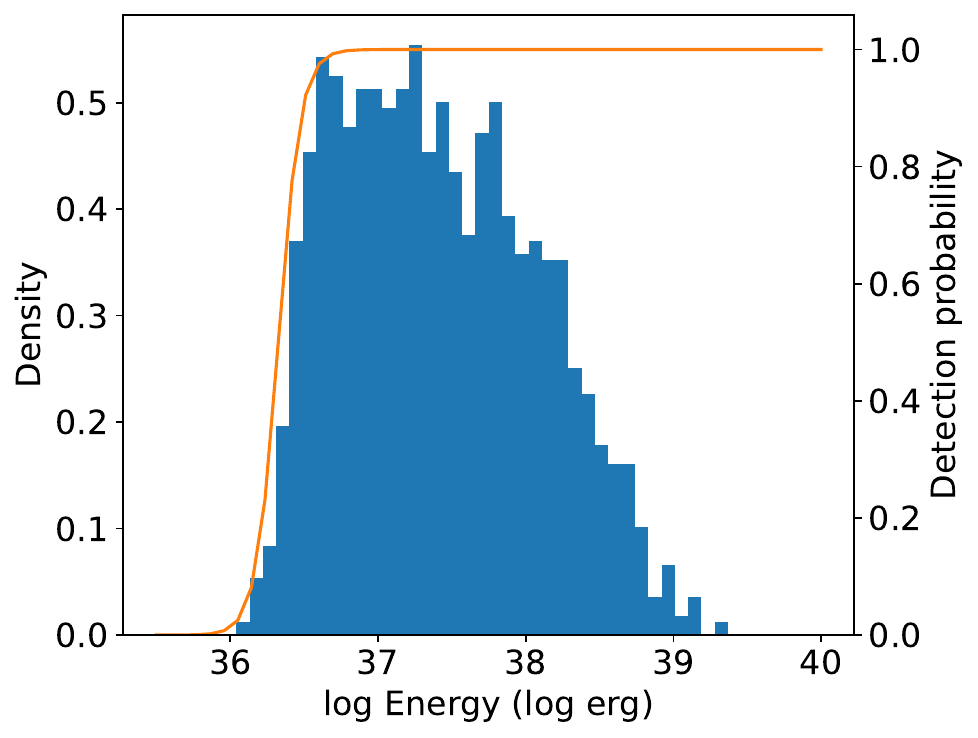}
    \label{fig:dect_1124AA}}
\subfloat[FRB 20201124A(B)]{
    \includegraphics[width=0.48\textwidth]{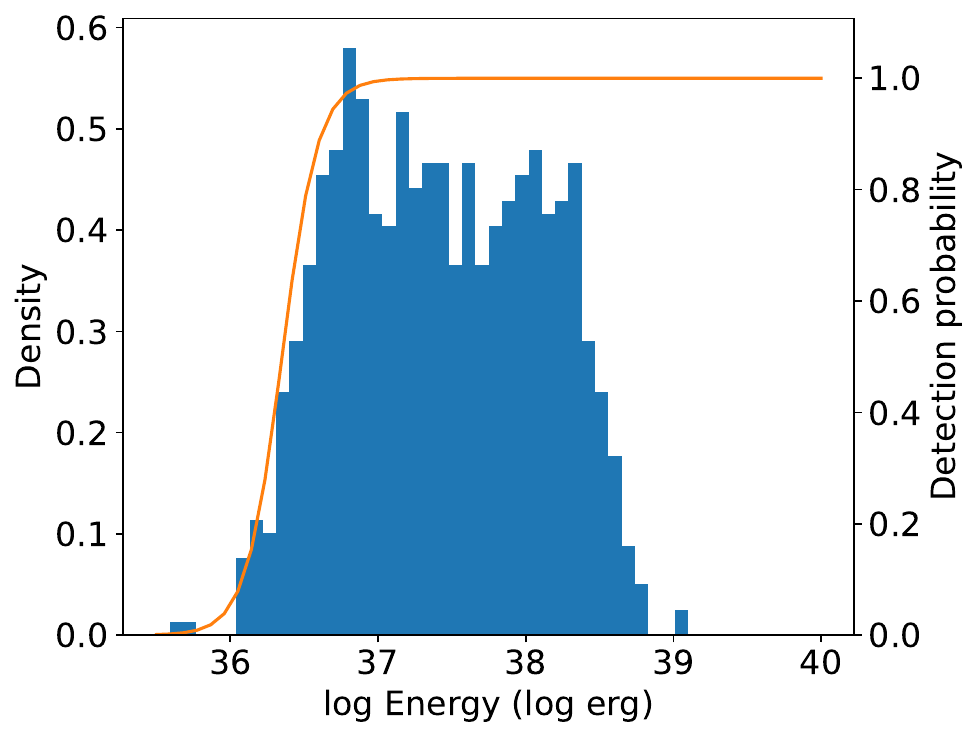}
    \label{fig:dect_1124AB}}\\
\subfloat[FRB 20121102A]{
    \includegraphics[width=0.48\textwidth]{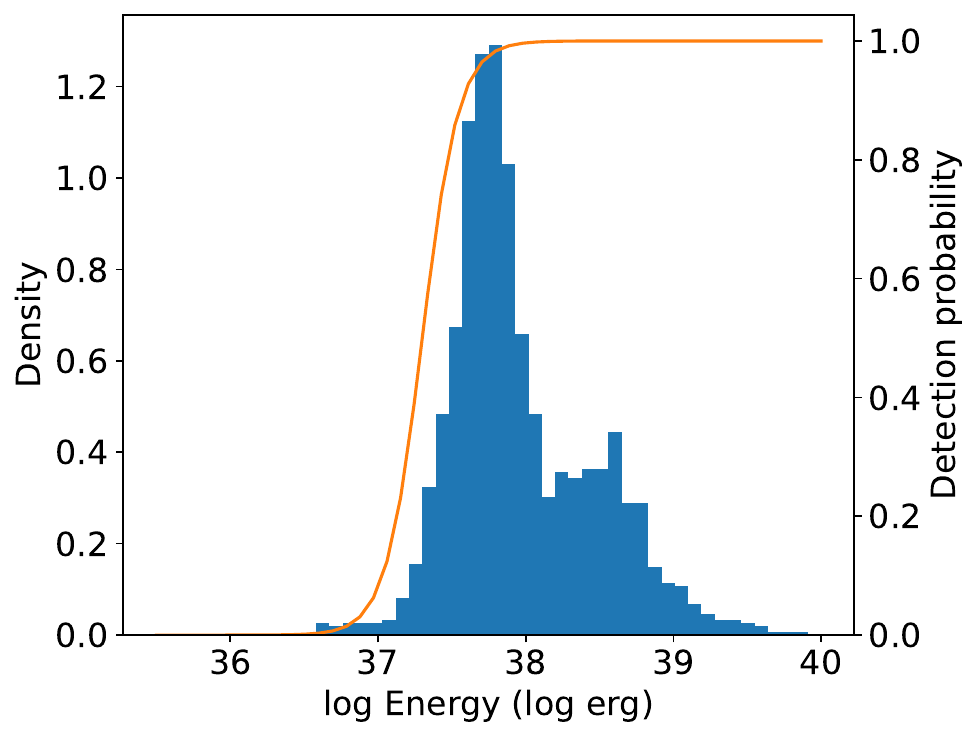}
    \label{fig:dect_121102}}
\subfloat[FRB 20190520B]{
    \includegraphics[width=0.48\textwidth]{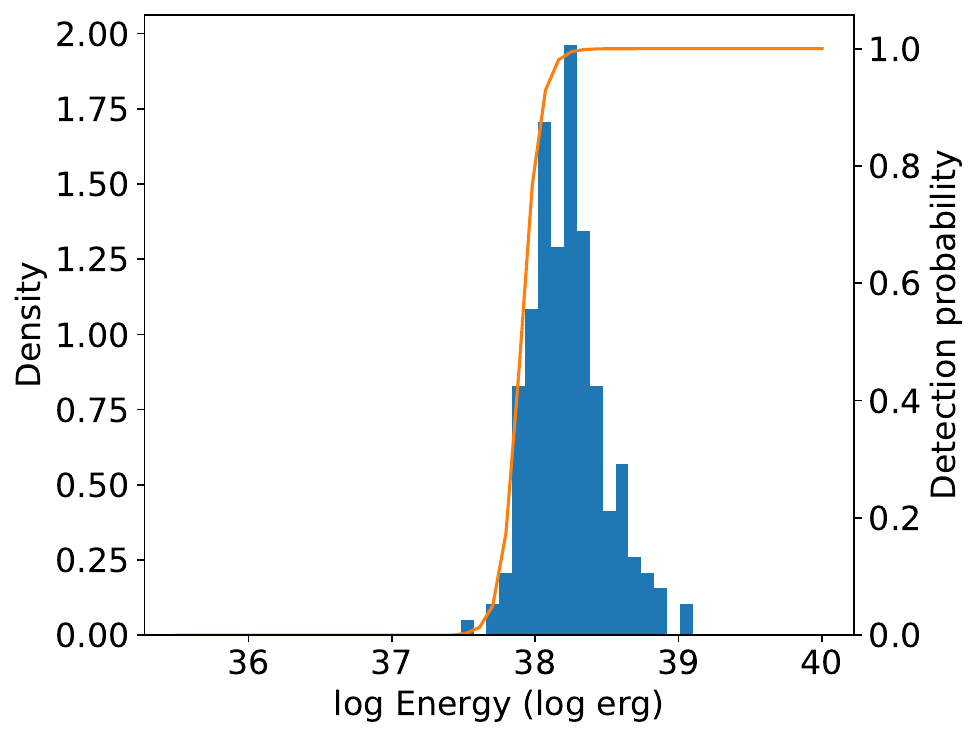}
    \label{fig:dect_190520}}\\
\subfloat[FRB 20220912A]{
    \includegraphics[width=0.48\textwidth]{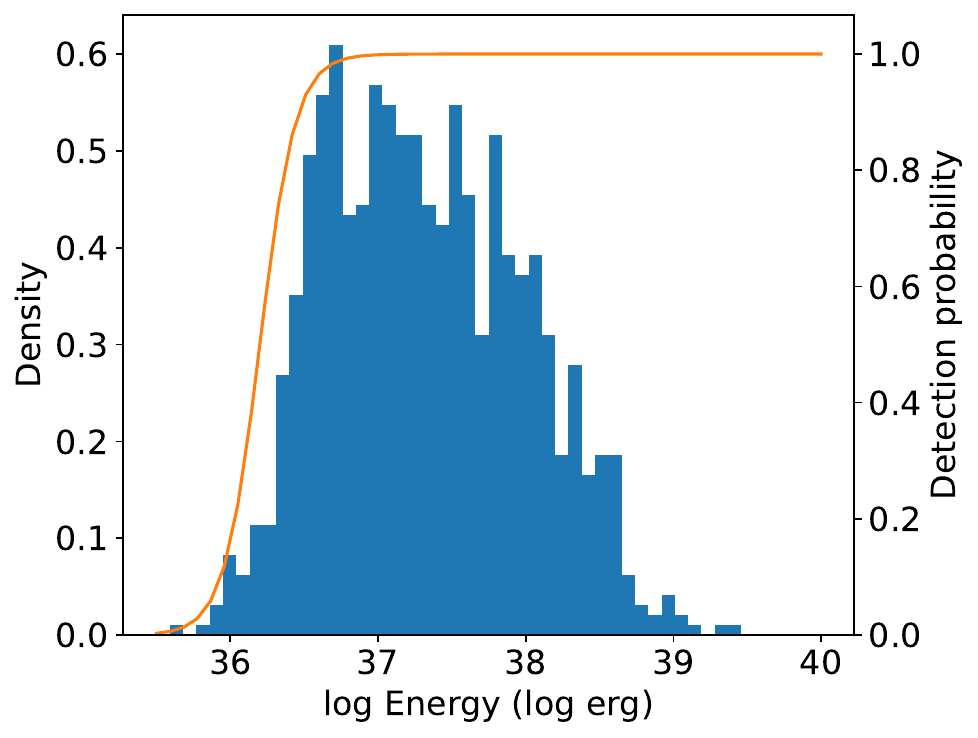}
    \label{fig:dect_220912}}
\subfloat[SGR J1935+2154]{
    \includegraphics[width=0.48\textwidth]{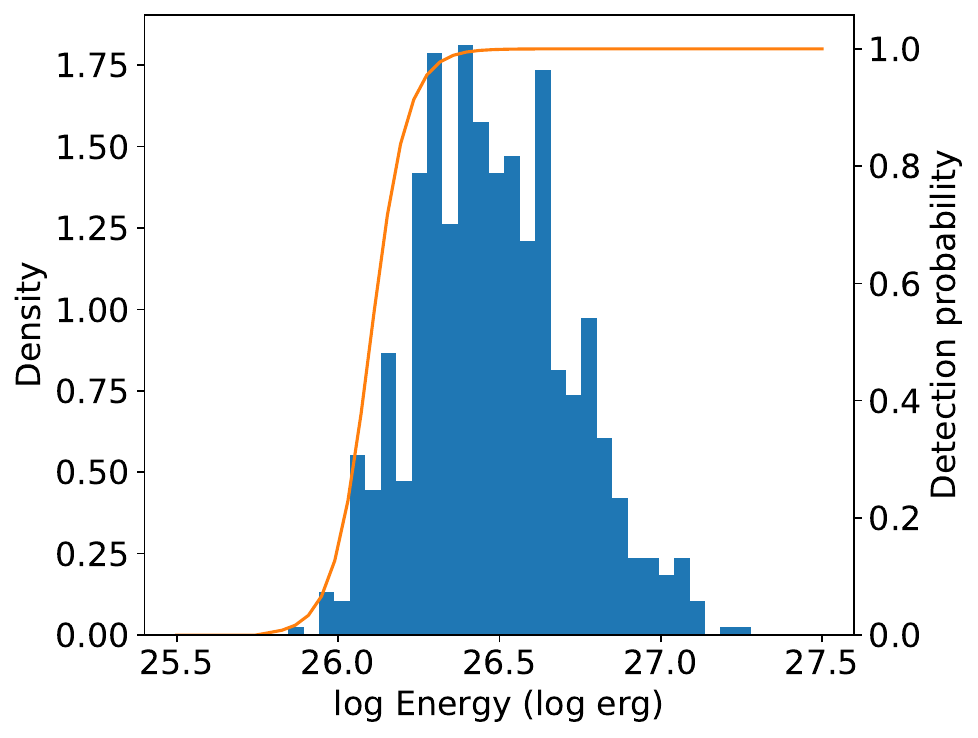}
    \label{fig:dect_sgr1935}}
\caption{Detection probabilities and energy distributions of different FRBs and the pulsar phase of SGR J1935+2154. The plots for the FRBs have the same energy range to facilitate comparison. Note that the detection probabilities are drawn on a different scale for readability.}
\label{fig:dect_prob}
\end{figure*}

\section{MCMC corner plots}
The following Figures shows the MCMC corner plots from the fitting for the FRBs and the pulsar phase of SGR J1935+2154.

\begin{figure*}[ht]
\centering
\includegraphics[width=0.99\textwidth]{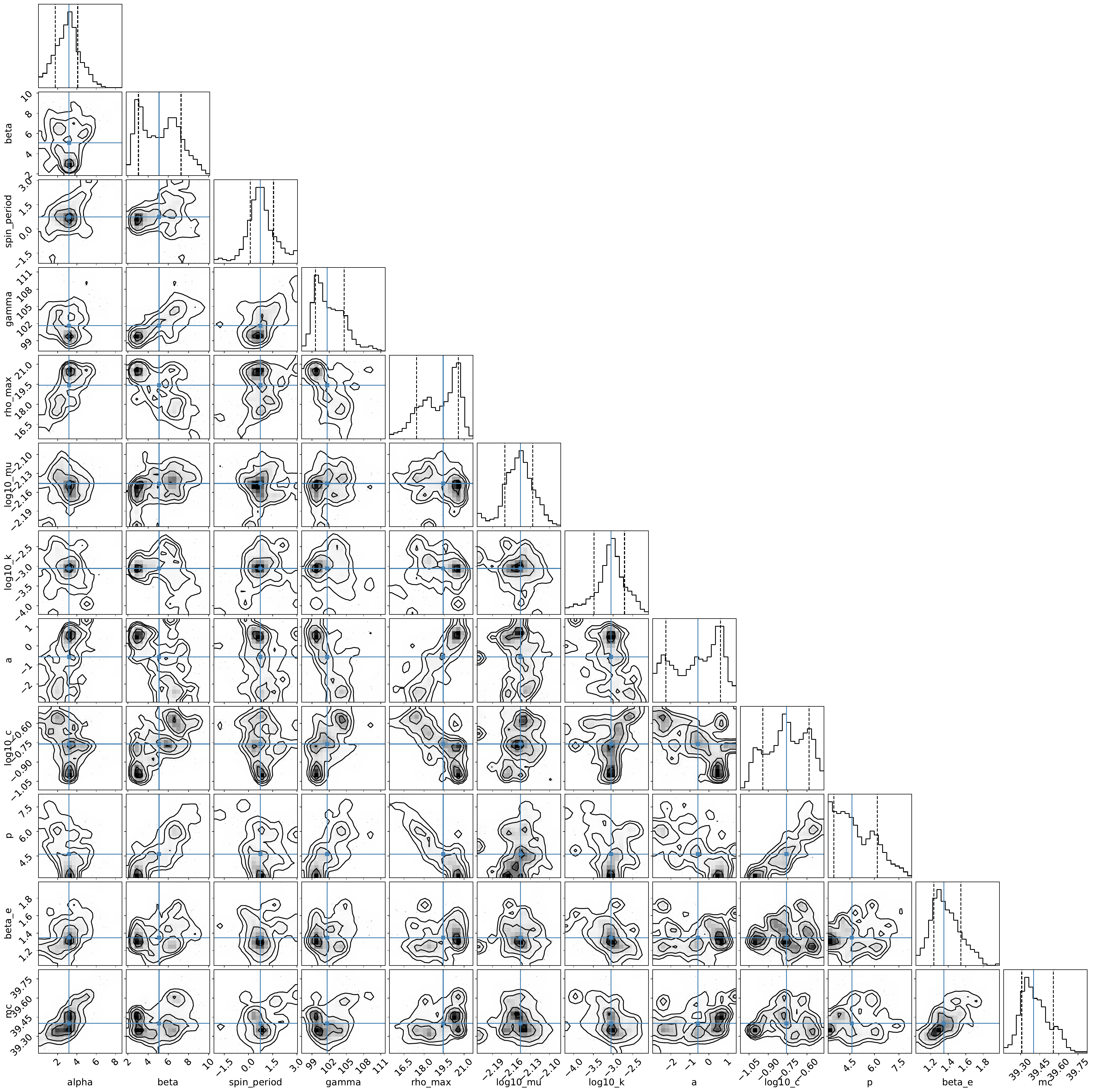}
\caption{MCMC corner plot for FRB 20201124A(A).}
\label{fig:mcmc_1124aa}
\end{figure*}

\begin{figure*}
\centering
\includegraphics[width=0.99\textwidth]{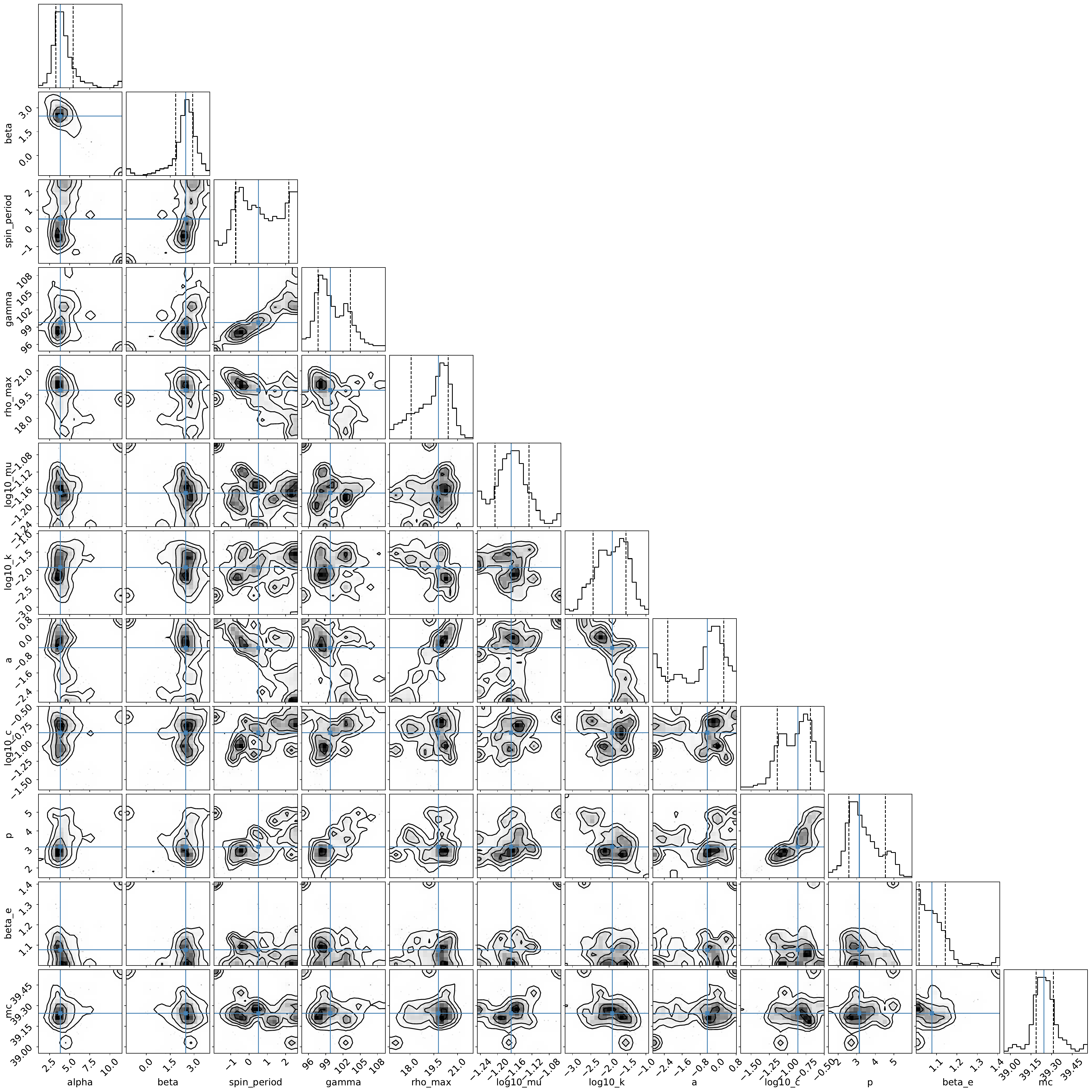}
\caption{MCMC corner plot for FRB 20201124A(B).}
\label{fig:mcmc_1124ab}
\end{figure*}

\begin{figure*}
\centering
\includegraphics[width=0.99\textwidth]{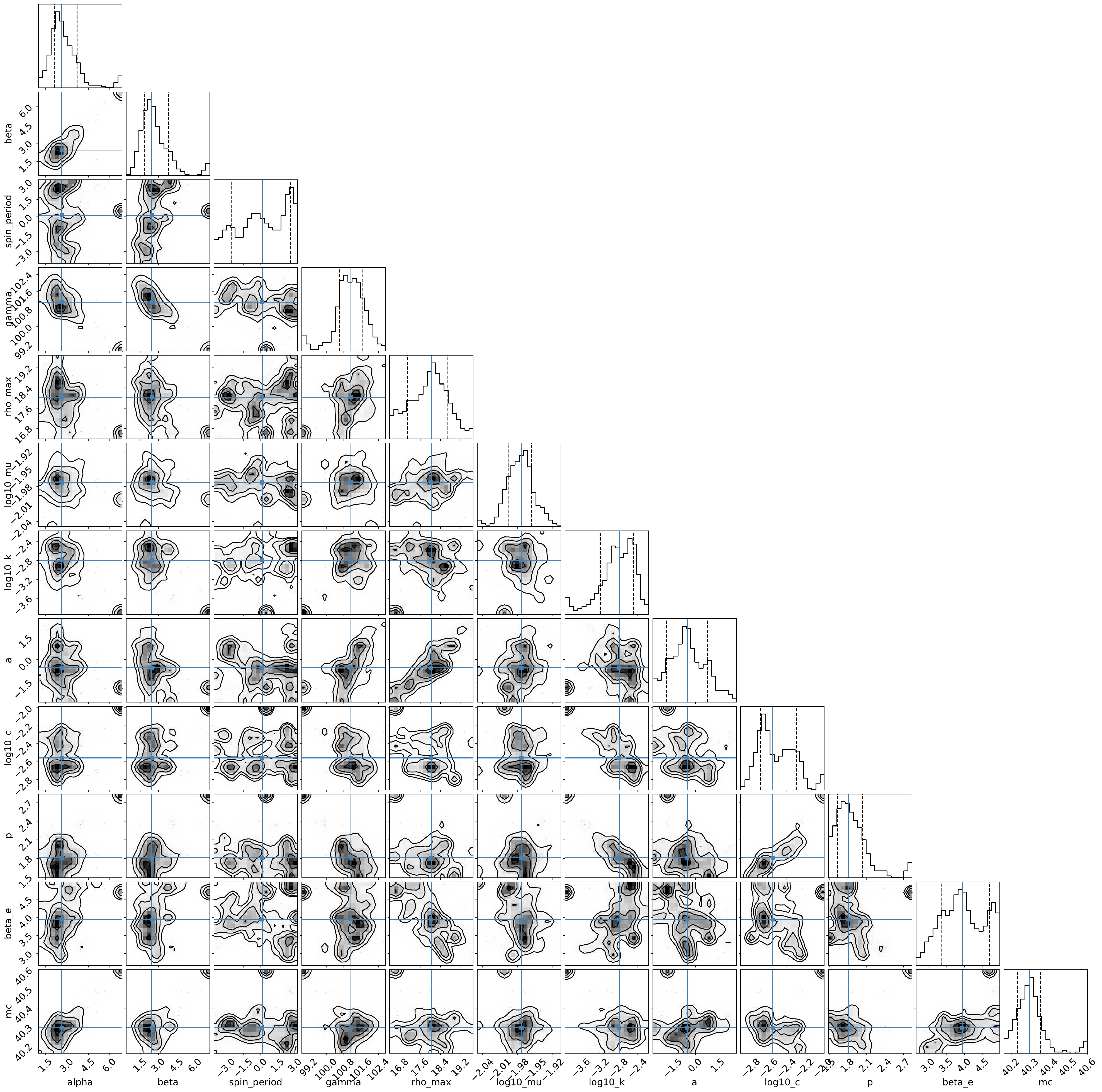}
\caption{MCMC corner plot for FRB 20121102A.}
\label{fig:mcmc_121102}
\end{figure*}

\begin{figure*}
\centering
\includegraphics[width=0.99\textwidth]{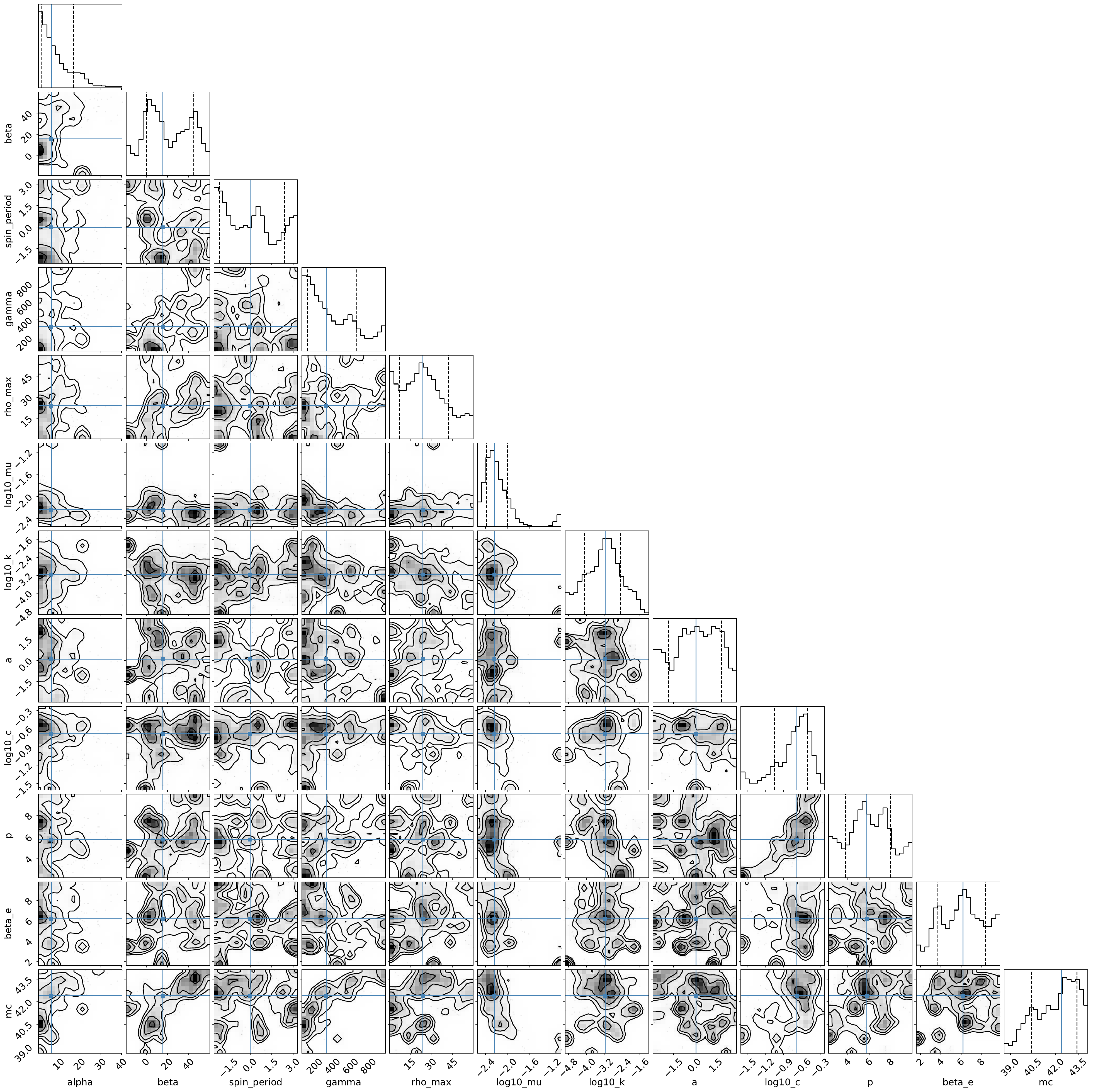}
\caption{MCMC corner plot for FRB 20190520B.}
\label{fig:mcmc_190520}
\end{figure*}

\begin{figure*}
\centering
\includegraphics[width=0.99\textwidth]{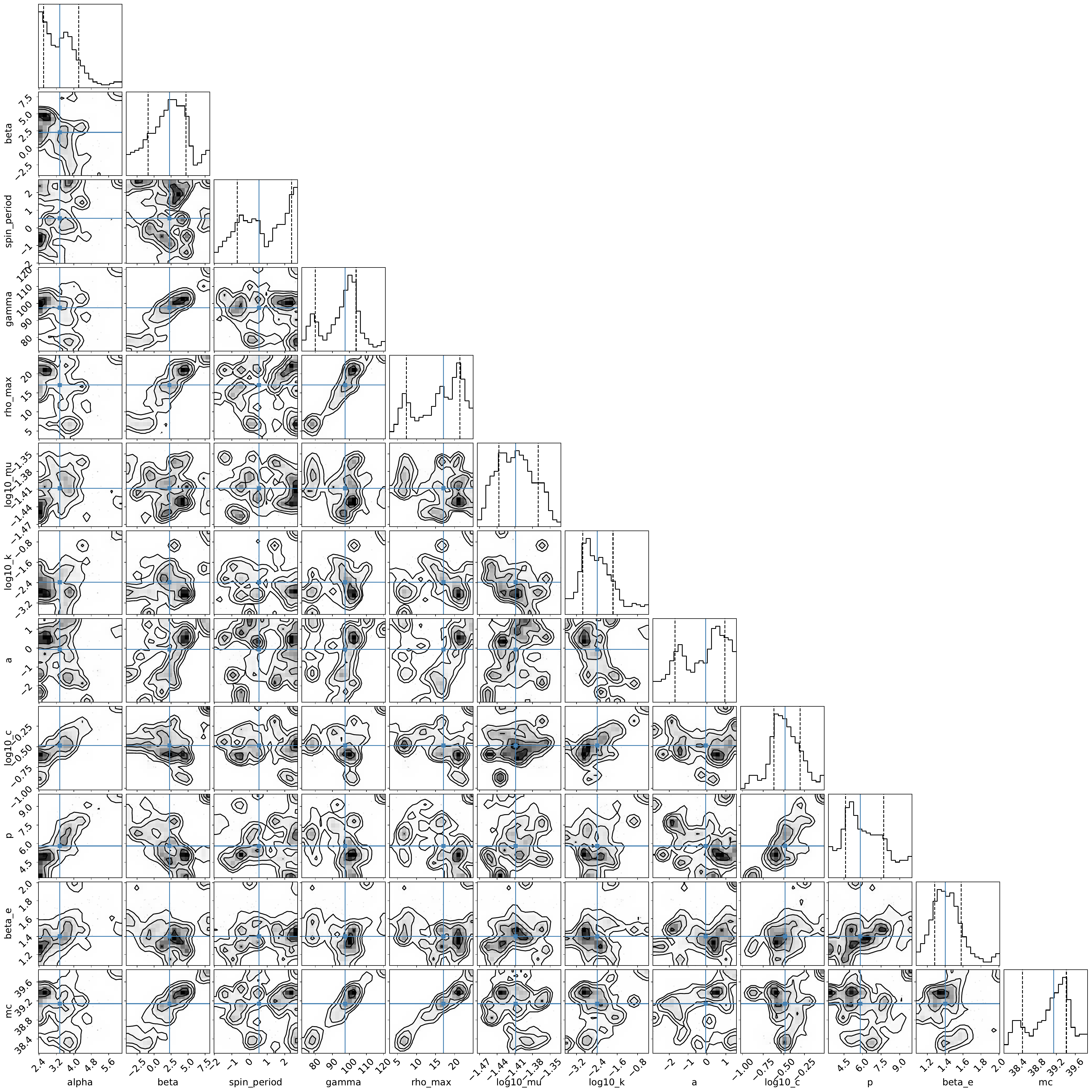}
\caption{MCMC corner plot for FRB 20220912A.}
\label{fig:mcmc_220912}
\end{figure*}

\begin{figure*}
\centering
\includegraphics[width=0.99\textwidth]{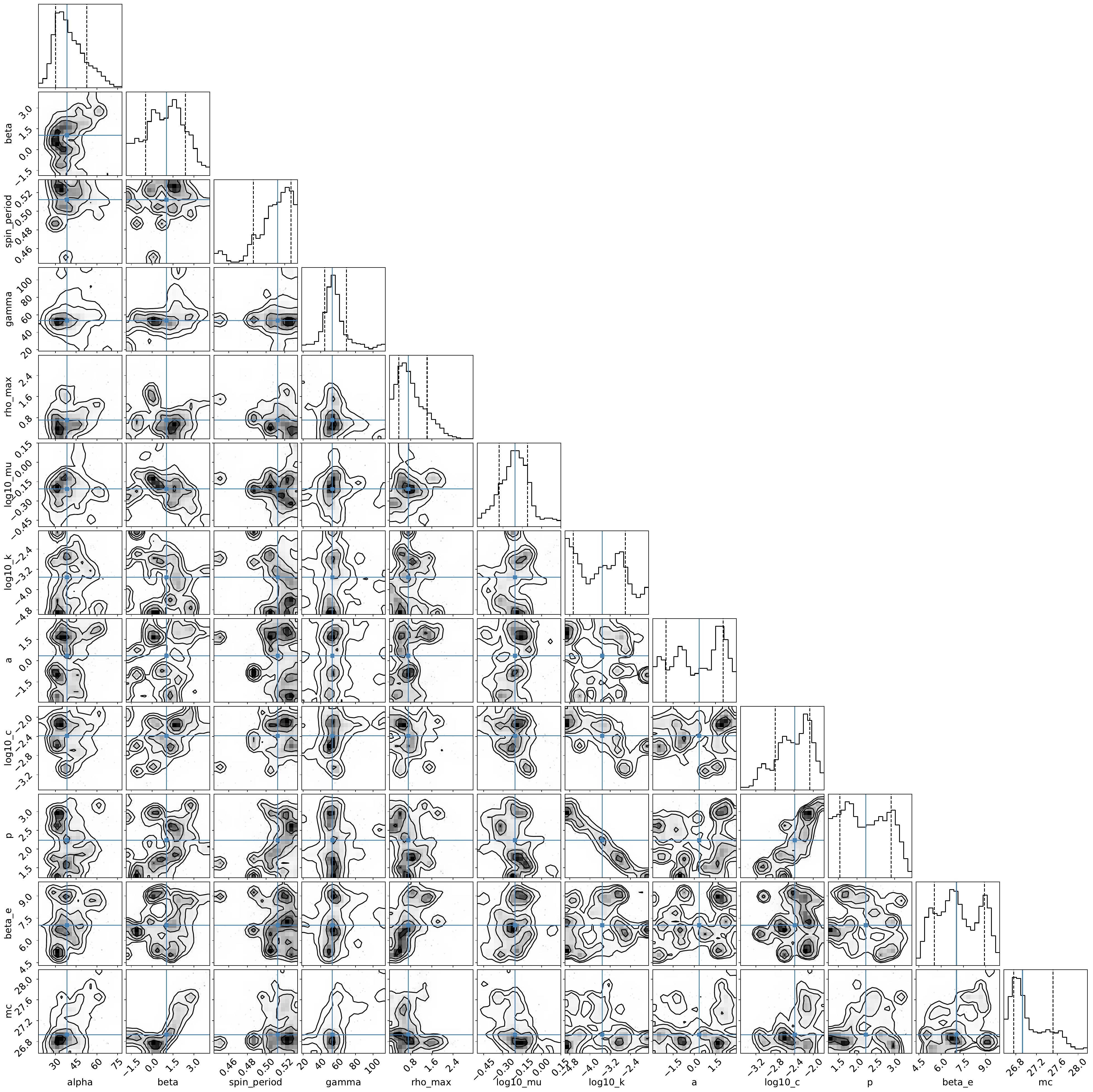}
\caption{MCMC corner plot for SGR J1935+2154.}
\label{fig:mcmc_sgr1935}
\end{figure*}
\end{document}